\documentclass[aps,prb,twocolumn,letterpaper]{revtex4}
\usepackage{graphicx}
\usepackage{dcolumn}
\usepackage{amsmath}
\usepackage{amssymb}
\usepackage{float}
\usepackage{epstopdf}
\usepackage{color}

\def\rv{{\bf r}}

\def\kv{{\bf k}}
\def\qv{{\bf q}}

\def\beq{\begin{equation}}
\def\eeq{\end{equation}}
\def\beqa{\begin{eqnarray}}
\def\eeqa{\end{eqnarray}}

\begin{document}

\title{Quasiparticle Interference Patterns in a Topological Superconductor}
\author{Aaron Farrell$^1$, Maxime Beaudry$^{1,2}$, M. Franz$^3$ and T. Pereg-Barnea$^1$}
\affiliation{$^1$Department of Physics and Centre for the Physics of Materials, McGill University,
Montreal, QC, Canada\\
$^2$D\'epartement d'Informatique, Universiti\'e de Montr\'eal, Montreal, QC, Canada \\
$^3$Department of Physics and Astronomy, University of British Columbia,
Vancouver, British Columbia, Canada}
\date{\today}
\begin{abstract}
In light of recent proposals to realize a topological superconductor on the surface of strong topological insulators, we study impurity and vortex scattering in two dimensional topological superconductivity.
We develop a theory of quasiparticle interference in a model of the surface of a three dimensional strong topological insulator with a pairing term added. We consider a variety of different scatterers, including magnetic and nonmagnetic impurity as well as a local pairing order parameter suppression associated with the presence of a vortex core.
Similar to the case of a surface of a three dimensional topological insulator without pairing, our results for non-magnetic impurity can be explained by the absence of back scattering, as expected for a Dirac cone structure.
In the superconducting case, doping away from the Dirac point leads to a doubling of the contours of constant energy. This is in contrast to the unpaired case where the chemical potential simply adds to the bias voltage and shifts the energy.  This doubling of contours results in multiplying the number of possible scattering processes in each energy.  Interestingly, we find that some processes are dominant in the impurity case while others are dominant in the vortex case.  Moreover, the two types of processes lead to a different dependence on the chemical potential.
\end{abstract}
\maketitle

\section{Introduction}

The search for a Majorana mode in a condensed matter system is currently a very active area of research.
Some very promising experimental  results have been reported in one dimension\cite{Experiment1,Experiment2} while in two dimensions appealing theoretical proposals exist\cite{Fu, Tanaka}.  The above systems are based on the combination of momentum spin locking with superconducting pairing.  In two dimensions this combination is provided by heterostructures of a superconducting layer in contact with either a three dimensional topological insulator (3DSTI)\cite{Fu} or a spin-orbit coupled semiconductor\cite{Sau, Alicea}.  While the superconductor furnishes the pairing, the spin orbit coupled layer provides the momentum-spin locking\cite{Koren,Sacepe,Qu,Williams,Cho,Xu}.  Another type of proposal makes use of an innate tendency for developing superconductivity in materials with spin orbit coupling (SOC)\cite{Zhao,Farrell1,Farrell2,Farrell3}.

Given the above, it is timely to search for unique properties of a system where superconductivity arises from an underlying Dirac-like band structure. In this Paper we develop a theory of quasiparticle interference (QPI)  patterns in such systems. The QPI patterns are of direct experimental relevance as they can be measured using the technique of Fourier-transform scanning tunnelling spectroscopy (FT-STS)\cite{Hoffman,He,Vershinin}.
This method measures the local density of states (LDOS) of a sample in the vicinity of a single impurity. Theoretically, the pattern which is observed is dictated by the underlying clean system ({\it i.e} without the impurity) and so properties of the clean system can be deduced from such a pattern and readily compared with calculations\cite{GuoFranz, TamiQPI1,TamiQPI2,TamiQPI3}.

We consider the surface of a strong topological insulator with pairing added. We do this by treating two complementary models. The first is a single Dirac cone in the continuum and should capture universal properties of a Dirac band structure in the presence of pairing. Second, we scrutinize our continuum model results by looking at a more physical lattice model. This model has been developed\cite{MarchandFranz} for the surface of a strong topological insulator. It specifically avoids the doubling theorem by including both surfaces of the material. We calculate the QPI pattern for a variety of scatterers, including a charge defect, a magnetic defect and a defect in the superconducting order parameter.

\begin{figure}[b] \label{fig0}
  \setlength{\unitlength}{.5mm}
  \includegraphics[width = 8.5cm]{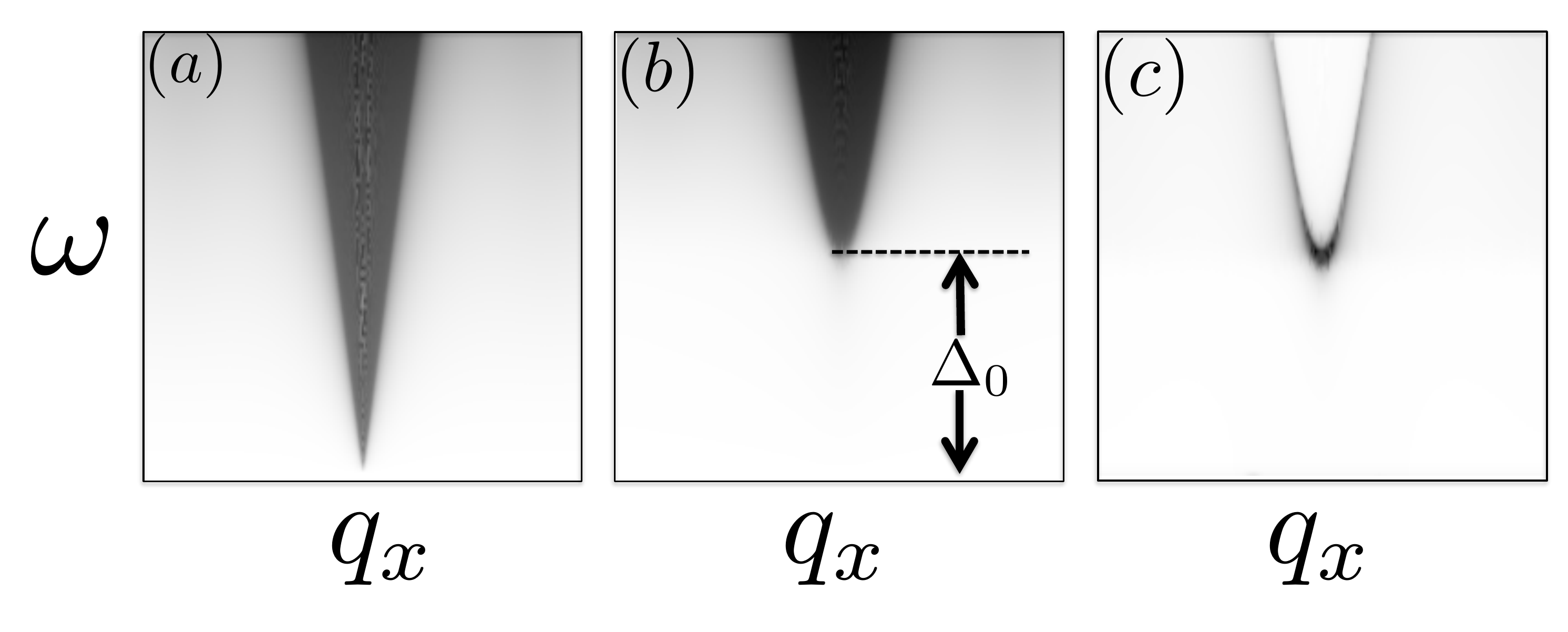}
\caption{Quasiparticle interference patters, observable by FT-STS, on the surface of a strong topological insulator with SC order. Panels a) and b) show the effect of non-magnetic impurities on the normal ($\Delta_0=0$) and SC state, respectively. Panel c) displays the effect of disorder in the order parameter amplitude.
}
\end{figure}
The QPI for a strong topological insulator in the presence of charge and magnetic impurities has been calculated by one of us in the past\cite{GuoFranz}. With the chemical potential tuned to the Dirac point, we obtain results consistent with this previous study with the exception that  QPI pattern is only observable for energies above the superconducting gap $\Delta_0$, as illustrated in Fig.\ \ref{fig0}. As noted previously, unlike in normal metals where QPI shows distinctive peaks, patterns in STIs with non-magnetic impurities are non-singular and exhibit only an edge. We find this behavior to persist when the SC order is included.  An interesting difference however occurs when we consider disorder in the SC order parameter  amplitude $\Delta$ which will generically arise in the presence or pair-breaking impurities of vortices. In this case, QPIs display a sharp peak, as illustrated in Fig.\ \ref{fig0}c.

Moving the chemical potential away from the Dirac point does not result in any angular dependence of the properties of this pattern. However, we do see some interesting behavior in these patterns. We find a contour of singular LDOS features. As the chemical potential is increased (decreased) with respect to the Dirac point, the radius of the contour in the impurity induced QPI patterns is also increased (decreased) and changes linearly with the chemical potential. On the other hand, in the order parameter suppression QPI pattern, the radius of the singular contour is {\em independent} of the chemical potential. We trace this difference back to the fact that different quasiparticle scattering processes are favored in these two situations. It can also be argued that this particular dependence on $\mu$ for these two types of scatterers comes from the underlying Dirac band structure. These observations in the continuum model can be understood heuristically via chirality angle arguments and are supported by our lattice model calculations.

Although tuning the chemical potential in these systems may be challenging from an experimental point of view, the underlying result ({\it e.g.} the behavior of the QPI singularity) can be observed {\em via} other means. As an example, we suggest a simplified scenario where the sample we consider is placed in a capacitor. This adds a bias voltage between the edges of the material. The voltage acts as a chemical potential with differing signs on each edge of the sample. By changing the bias one can observe the distinct behavior the QPI due to an impurity or a vortex. There has recently been an experimental realization of a set-up analogous to this capacitor system\cite{ChangChemicalPotential}. In this work it was shown that the chemical potential on each edge of thin film 3DSTI can be tuned independently of the other edge through the use of dual-gate structures. Our results are then directly applicable to an FT-STS measurements on such a device when the chemical potential on one edge is tuned to be opposite to each other edge.

The rest of this paper is organized as follows. In the next section we give an overview of the methods we have followed to obtain our results. We present our method for calculating the QPI pattern, define each type of perturbation, present our two model systems and discuss numerical details. Section \ref{sec:Results} then presents our results for the QPI patterns. We begin by tuning the chemical potential to the Dirac point, where the QPI pattern for our continuum model can be calculated exactly (the details of this are relegated to the Appendix). This not only helps us compare our results with a previous study, but also allows us to develop some intuition. We then move on to the main result of this paper, the behavior of QPI patterns at finite chemical potential. Here we present numerical data as well as further discussion. We then discuss an alternative to tuning the chemical potential and perform an explicit calculation using our lattice model.

\section{Methodology}
\subsection{Green's function for a single impurity and the Born approximation}
We begin by discussing a general expression for the total Green's function. The approach taken here will be similar to previous work\cite{GuoFranz}, but appropriately generalized to include superconductivity and to be suitable for application to our lattice model. In either case we take a total Hamiltonian $H = H_0 + H_{imp}$, the first term describes the underlying clean system while the latter term describes the perturbation.  The clean system's Hamiltonian is given by:
\beq
H_0 = \frac{1}{2} \sum_{\kv} \psi_{\kv} ^\dagger \mathcal{H}_{\kv} \psi_{\kv}
\eeq
here we have defined $\psi_{\kv} = (c_{\kv,\alpha}, c_{-\kv, \alpha}^\dagger)^T$ where $\alpha$ labels additional degrees of freedom in the model (spin, sub lattice, etc.).  The matrix $\mathcal{H}_{\kv}$ is given by
\beq
\mathcal{H}_{\kv} = \left(   \begin{matrix} 
      h(\kv) & \Delta(\kv) \\
       \Delta^\dagger(\kv)  & -h^\dagger(-\kv) \\
   \end{matrix}\right)
\eeq
where $\Delta_{\kv}$ and $h(\kv)$ are matrices in the space of the degrees of freedom implicit in $\alpha$ before, $h(\kv)$ describes the normal band structure of the model and $\Delta(\kv)$ describes the pairing. Next we consider a local perturbation which couples to the electronic degrees of freedom in some general form:
\beq
H_{imp} = \frac{1}{2N} \sum_{\kv,\kv'} \psi_{\kv}^\dagger V^0_{imp}(\kv-\kv') \psi_{\kv'}
\eeq
where $N$ is the number of lattice sites; $V^0_{imp}(\kv-\kv')$ is the Fourier transform of the perturbation potential.  Its matrix structure depends on the impurity we consider and will be discussed in detail below.

Next we consider calculating the Green's function for this model. We define the Green's function as follows
\beq
G(\kv,\kv',\tau) = -\langle \text{T}_{\tau}\left[ \psi_{\kv} (\tau) \psi^\dagger_{\kv'}(0) \right]\rangle
\eeq
where $\tau$ is the imaginary time and $\psi_{\kv}(\tau)$ is the Heisenberg picture version of $\psi_{\kv}$. Fourier transforming to Matsubara frequency and defining the $T$-matrix we write
\beqa
G(\kv,\kv',i\omega_n)& =&  G^0(\kv,i\omega_n) \delta_{\kv,\kv'}   \\ \nonumber &+&  G^0(\kv,i\omega_n)   T(i\omega_m,\kv,\kv')G^0(\kv',i\omega_n)
\eeqa
where $G^0(\kv,i\omega_n)=\left(i\omega_m-\cal{H}_\kv\right)^{-1}$ is the Green's function of the clean system.  In the case of a local perturbation, where the potential is a $\delta$-function in real space, the potential and the $T$-matrix are momentum independent and the self consistency equation for the $T$-matrix can be solved:
 \beq
 T(i\omega_m) = N^{-1}V^0_{imp}\left(I - N^{-1}\sum_{\kv} G^0(\kv,i\omega_n)   V^0_{imp} \right)^{-1}.
 \eeq
In principle poles in the $T$-matrix reveal information about bound states in the system. However, in this study we are interested in the QPI patterns in momentum space.  Since the $T$-matrix has no momentum dependence, the results of a weak perturbations will not differ from the simpler Born approximation.
%
To this end, we keep only the leading order term in $V^{0}_{imp}$ which leaves
\beqa\label{Greenfunction}
G(\kv,\kv',i\omega_n) &\simeq&  G^0(\kv,i\omega_n) \delta_{\kv,\kv'}   \\\nonumber &+& \frac{1}{N} G^0(\kv,i\omega_n)  V^0_{imp}(\kv-\kv') G^0(\kv',i\omega_n).
\eeqa
Moreover, in the Born approximation, which is appropriate away from impurity bound state energies, the momentum dependence of the potential separates from that of the Greens functions convolution when calculating the Fourier transformed density of states\cite{Capriotti}.

\subsection{Local density of states and impurity types}

Here we will describe how we calculate the various LDOS patterns in the work to come. We are considering FT-STS experiments where the tip of the STM is normal (not magnetic or superconducting) and measures the LDOS, theoretically this quantity is given as follows
\beq
n(\omega, \rv)= -\frac{1}{2\pi} \text{Im} \left(\text{Tr} \left[(1+\tau^z)G(\rv,\rv, \omega+i\eta)\right]\right)
\eeq
where $\tau^z$ acts on particle-hole degrees of freedom, $\eta$ is an infinitesimal and $G(\rv,\rv', \omega+i\eta)$ is the double Fourier transform of  $G(\kv,\kv',\omega+i\eta)$ defined in Eq. (\ref{Greenfunction}) above.
Fourier transforming one finds that the modulation to the local density of states in the Born approximation is
\begin{widetext}
\begin{eqnarray}\label{LDOScharge}
\delta n(\omega, \qv) &=& -\frac{1}{2\pi N}  \sum_{\kv} {\text Im} \left(\text{Tr} \left[(1+\tau^z)G^0(\kv, \omega+i\eta)  V^0_{imp}  G^0(\kv+\qv,\omega+i\eta)    \right]\right)
\end{eqnarray}
\end{widetext}
where ${\text Im}(f(w+i\eta))\equiv(f(w+i\eta)-f(w-i\eta))/2i$.

Following Ref. [\onlinecite{GuoFranz}], we are interested in a more general interference pattern. Here we imagine that the STM tip can resolve different degrees of freedom ({\it e.g.} spin can be resolved by a spin polarized tip).
It is therefore possible to insert a general matrix $V_\alpha$ into the trace.  This matrix acts to resolve the component of the LDOS pattern of interest\cite{GuoFranz}.

We now discuss the various types of perturbations that will be considered. Here we will focus on a physical description of these impurities and leave the formal details to the Appendix. We begin with a simple {\em charge} impurity. Here we imagine that the chemical potential at a certain site has been altered. Second, we are interested in magnetic impurities. These alter the local Zeeman splitting on a single site and so couple to the spin of the electrons.  In what follows we will combine charge and magnetic impurities into a single heading which we will refer to as impurities. It is then useful to define the QPI patterns for any of these impurities as
\begin{widetext}
\begin{eqnarray}\label{LDOSparticle}
\delta n_{\alpha,\beta}(\omega, \qv) &=& -\frac{1}{2\pi N}  \sum_{\kv} {\cal I} \left(\text{Tr} \left[V^{\alpha}(1+\tau^z)G^0(\kv, \omega+i\eta) V^{\beta} G^0(\kv+\qv,\omega+i\eta)    \right]\right).
\end{eqnarray}
\end{widetext}
The matrices $V^{\beta}$ are outlined in the appendix. The first label, $\alpha$, denotes the type of STM tip. $\alpha=0$ is a normal (charge) tip while $\alpha=1,2$ or $3$ resolves the component of the electron's spin along the $x,y$ or $z$ direction respectively. Meanwhile, $\beta$ labels the type of impurity we are considering. $\beta=0$ is a charge impurity while $\beta=1,2$ or $3$ refers to a magnetic impurity with its spin along the $x,y$ or $z$ axis. When we refer to patterns such as $\delta n_{1,2}(\omega, \qv)$ it is the patterns above to which we are referring. It should be noted that any physical QPI pattern due to charge/spin scattering can be written as a combination of the above.

The second class of perturbation we are interested in is a local suppression in the superconducting order parameter. We will refer to this as OP suppression.  This perturbation can be thought of as a simplified description of a vortex where only the OP suppression at the vortex core is taken into account\cite{TamiQPI2}.  Alternatively, other types of disorder are sometimes accompanied by OP variations\cite{Lang}.  We refer to the LDOS in this case as $\delta n_{OP}$.

\subsection{Model Hamiltonians}

To discuss the surface of a 3DSTI we will make use of two model Hamiltonians.  The first is a continuum Dirac cone model which will be our primary focus.  This model is a single Dirac cone\cite{GuoFranz} with a chemical potential and $s$-wave pairing. Symbolically we have
\beq\label{continuum}
H_0 = H_{TI} + H_{SC}+H_\mu
\eeq
with
\beqa
&&H_{TI} = v \int d^2\kv c_{\kv}^\dagger (k_x\sigma_y - k_y\sigma_x)c_{\kv} \\ \nonumber
&&H_{\mu} = - \mu \sum_{\alpha} \int d^2\kv c_{\kv,\alpha}^\dagger c_{\kv, \alpha}\\ \nonumber
&&H_{SC} = \Delta_0  \int d^2\kv  \left( c_{\kv,\uparrow}^\dagger c_{-\kv, \downarrow}^\dagger + c_{-\kv, \downarrow}c_{\kv,\uparrow}\right)
\eeqa
where $c_{\kv}=(c_{\kv,\uparrow}, c_{\kv,\downarrow})^T$, $\mu$ is the chemical potential and $\Delta_0$ is the OP amplitude. We have adopted units such that $\hbar=1$. Such a model can be diagonalized and yields the two eigenvalues at each wave vector $E_{\pm}(\kv)=\sqrt{\Delta_0^2+(\pm v|\kv|-\mu)^2}$. Further, finding an exact, closed-form expression for the Green's function of the above system is possible. We have outlined the details this calculation in the Appendix.

The second model is a lattice model of a 3DSTI where only the surfaces of this system are considered (we will arbitrarily refer to the two surfaces as 'top' and 'bottom').  This model will be used sparingly and will mostly be employed as a physical consistency check for features we find in the continuum model.

In order to model the single Dirac cone on the surface of a 3DSTI we follow Marchand and Franz\cite{MarchandFranz}.  Their approach is briefly outlined here.  The doubling theorem states that in a time reversal invariant periodic systems Dirac points appear in pairs.  Therefore, any two dimensional lattice model can not have an odd number of Dirac points.  The 3DSTI is indeed a periodic system.  However, it avoids the doubling theorem by placing half of its Dirac points on each surface.  To mimic this we employ a two-surface model.  This is a minimal way to model a surface with an odd number of Dirac points.

In this approach, one begins with a three dimensional model and integrates out the bulk, leaving only the two edges of the material. Adopting this model gives our clean system's Hamiltonian:
\beq\label{lattice}
h(\kv) = \left(      \begin{matrix} 
      \tilde{h}_\kv & \tilde{M}_\kv&0&0 \\
      \tilde{M}_\kv & -\tilde{h}_\kv&2R_\kv&0 \\
      0 &2R_\kv^\dagger&\tilde{h}_\kv & \tilde{M}_\kv \\
      0&0&\tilde{M}_\kv & -\tilde{h}_\kv \\
   \end{matrix}\right)
\eeq
where $\tilde{h}_\kv=2\lambda(\sin{k_x}\sigma_y - \sin{k_y}\sigma_x)-\mu\sigma_0$ and $\tilde{M}_{\kv} = \epsilon-2t(\cos{k_x}+\cos{k_y})$ and $R_{\kv} = \frac{1}{4}\left(\epsilon+2t(\cos{k_x}+\cos{k_y})\right)$ where $\epsilon=4t$. The matrix above is an $8\times8$ matrix, the sub matrices $\tilde{h}_\kv, \tilde{M}_\kv$ and $R_{\kv}$ act on a $2\times2$ space of spin. The basis of the above is $[( \uparrow, 1, T), (\downarrow, 1, T), ( \uparrow, 2, T), (\downarrow, 2, T) , ( \uparrow, 1, B), (\downarrow, 1, B), ( \uparrow, 2, B), (\downarrow, 2, B)]$ where arrows refer to spin, the numbers refer to one of two bands and '$T$' and '$B$' refer to top and bottom edges. With this basis in mind, we see that  $\tilde{M}_\kv$ couples orbitals within the same surface and $R_{\kv}$ couples different surfaces.

We now introduce superconductivity pairing only within the same orbital and the same surface.
\beq
\Delta(\kv) = \left(      \begin{matrix} 
      \tilde{\Delta}_\kv & 0&0&0 \\
      0& \tilde{\Delta}_\kv&0&0 \\
      0 &0& \tilde{\Delta}_\kv& 0 \\
      0&0&0&\tilde{\Delta}_\kv \\
   \end{matrix}\right)
\eeq
where $\tilde{\Delta}_\kv = (i\sigma_y) \Delta_{\kv}$. We assume $s$-wave pairing and therefore take ${\Delta}_\kv=\Delta_0$. The above has 8 doubly degenerate eigenvalues:
\beq
E_\kv = \sqrt{\Delta_0^2+(\tilde{E}_{\kv}(s_1,s_2)-\mu)^2}
\eeq
where\cite{MarchandFranz} $\tilde{E}_{\kv}(s_1,s_2)=s_1\sqrt{\epsilon_\kv^2+(\sqrt{M_\kv^2+R_\kv^2}+s_2R_\kv)^2}$ with $s_1,s_2=\pm1$ and $\epsilon_\kv^2=4\lambda^2(\sin^2{k_x}+\sin^2{k_y})$. The clean Green's function of this system is not analytically tractable and so we rely on numerics for its calculation.

\subsection{Numerical method}

Here we briefly outline the numerical methods we use. We numerically compute the LDOS patterns for particle-like impurities as described in Eq.~(\ref{LDOSparticle}). The idea is to compute the Green's function in the absence of any impurity from the Hamiltonian using $G^0(\kv,i\omega_n)=\left(i\omega_m-\cal{H}_\kv\right)^{-1}$. Then, we produce the convolution with the impurity $V^{\beta}$. Once this is obtained, multiplying it with $V^{\alpha}$ then taking the trace produces the wanted results.

Each term in the perturbed Green's function is therefore a convolution of two functions of momentum, with each of these functions being an entry in the clean system's Green's function matrix.  In order to minimize run time and obtain high resolution LDOS maps we preform the convolution using the fast Fourier transform algorithm (FFT).  This amounts to first using FFT to express the two functions in real space, performing a direct product and then using FFT again to get back to momentum space.  This lowers the run time from $o(n^3)$ to $o(n\log n)$ where $n$ is the number of points in the Brillouin zone.

\section{Quasiparticle Interference Patterns in a Topological Superconductor}\label{sec:Results}
\subsection{Zero Chemical Potential}

\begin{figure*}[tb]
  \setlength{\unitlength}{.5mm}
  \includegraphics[scale=.5]{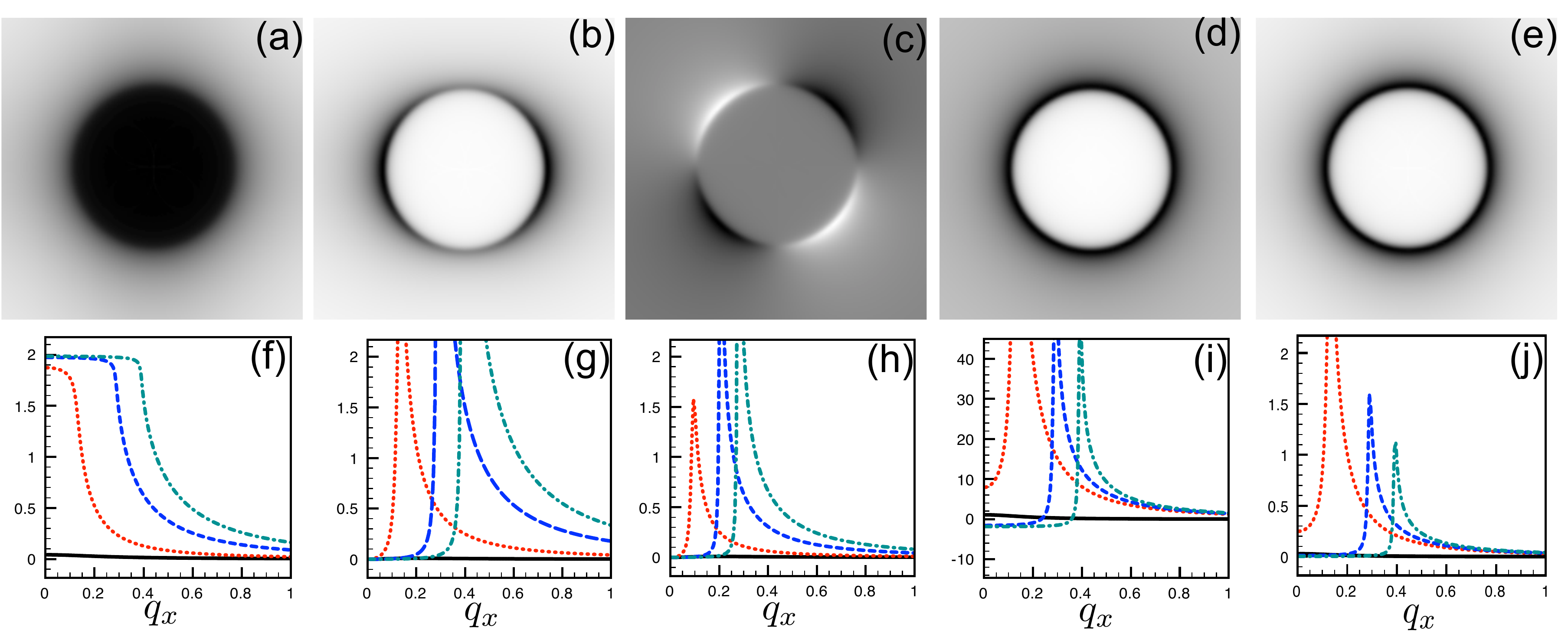}

\caption{{\small
Plots of the LDOS for both the lattice and continuum model with $\mu=0$. Along the top we have subfigures (a) through (e) for $\delta n_{00}$, $\delta n_{11}$, $\delta n_{12}$, $\delta n_{33}$, and $\delta n_{OP}$ (respectively) for the lattice problem. These calculations were done on a $600\times600$ lattice with $\lambda=1$, $t=0.5$, $\epsilon=2$, $\Delta_0=0.4$ and $\omega=0.6$. On the bottom subfigures (f) through (j) show plots of the expressions for  $\delta n_{00}$, $\delta n_{11}$, $\delta n_{12}$, $\delta n_{33}$, and $\delta n_{OP}$ (respectively) in the Dirac model. We have also explored the dependence of these patterns on the frequency $\omega$. We have fixed $\Delta_0=0.4$ and $v=1$ and plotted $\omega=0.385$ as a solid line (black online), $\omega=0.405$ as a dotted line (red online), $\omega=0.425$  as a dashed line (blue online) and $\omega=0.445$ as a dashed-dot line (teal online). With the exception of $\delta n_{12}$, which is plotted along the line $q_x=q_y$, all plots are along the line $q_y=0$.
     }
     }\label{fig:LDOSmuzero}
\end{figure*}

At the particle-hole symmetric point, i.e., $\mu=0$, the spectrum possesses Lorentz invariance.  This enables us to find a closed form solution to the continuum model, for all of the perturbations we consider.
This is done using the exact Green's function calculated in the appendix along with a standard Feynman parameterizations trick (see {\it e.g} [\onlinecite{peskin}]) which we will not repeat here. For the impurities these patterns  are most easily presented by defining the matrix $\mathcal{A}_{\alpha,\beta} (i\omega_m, \qv, \Delta_0)  = -\frac{1}{2}\delta n_{\alpha,\beta} (i\omega_m, \qv) $, the variables $z_1=2i\omega_m/q$ and $z_2=2\Delta_0/q$ and $z=\sqrt{-z_1^2+z_2^2}$ and the functions
\begin{eqnarray}
F(z) &=&2\sqrt{-1-z^2} \arctan\left[\frac{1}{\sqrt{-1-z^2}}\right],\\
G(z) &=&2\frac{\arctan\left[\frac{1}{\sqrt{-1-z^2}}\right]}{\sqrt{-1-z^2} }
\end{eqnarray}
and we have set $v=1$ hereafter for simplicity.  With these definitions we find
\begin{widetext}
\beq\label{muzero}
\mathcal{A}_{\alpha,\beta} (i\omega_m, \qv, \Delta_0)  = \small{\left(   \begin{matrix} 
      \log\left[1+\frac{\Omega^2}{\Delta_0^2-(i\omega_m)^2}\right]  -  F(z) & 0&0&0 \\
      0 &\hat{q}_x^2(2 +z^2 G(z))-1& \hat{q}_x\hat{q_y} \left(1+z^2G(z)\right) &-i \hat{q}_x z_1G(z)  \\
           0 & \hat{q}_x\hat{q_y} \left(1+z^2G(z)\right)& \hat{q}_y^2(2 +z^2 G(z))-1 &-i\hat{q}_y z_1G(z)  \\
                0 & i \hat{q}_x z_1 G(z) & i\hat{q}_y z_1G(z) &  - \log\left[1+\frac{\Omega^2}{\Delta_0^2-(i\omega_m)^2}\right]-(1+z_2^2)G(z)  \\
   \end{matrix}\right)}
\eeq
\end{widetext}
where $\Omega$ is an ultraviolet cut-off and $\hat{q}_i=q_i/|\qv|$. We note that with some small exceptions the above results are almost identical to those for a strong topological insulator {\em without} superconductivity\cite{GuoFranz} with  the replacement $(i\omega_m)^2\to (i\omega_m)^2-\Delta_0^2$.

The OP suppression can also be calculated in closed form at $\mu=0$ and yields the results
\begin{eqnarray}
\delta n_{\text{OP}}(i\omega_m, \qv)  &=& \frac{ 2\Delta_0 (i\omega_m)  G(z)  }{\pi ^2 q^2}   \\ \nonumber
\end{eqnarray}
For simplicity we have presented to above results in Matsubara space, the proper analytic continuation must be employed to obtain the physical QPI patterns.

We have compared the exact results above with those we have obtained from the numerical calculation at $\mu=0$ using the model in Eq.~(\ref{lattice}) with superconductivity added. We find that the two patterns agree remarkably well showing similar angular and radial features (the specifics of these features will be discussed below). We have showcased five of these patterns in density plots in the top of Fig.~\ref{fig:LDOSmuzero}. We now move on to discuss three properties of the QPI patterns we have found in the continuum model, keeping in mind the agreement between the lattice and continuum results.

First, let us discuss the radial features of the QPI patterns. Inspecting the above results we notice that the function $G(z)$ is singular at a $\qv$ length of $q=2\sqrt{\omega^2-\Delta_0^2}$ while $F(z)$ has a kink at this value (its derivative should be singular). To understand this value let us recall that the bulk energy bands for the continuum model are given by $E_{\kv,\pm} = \sqrt{\Delta_0^2+(\pm  |\kv|-\mu)^2}$. With $\mu=0$ the two bands become degenerate and contours of constant energy for a given frequency are then given by $|\kv|=\sqrt{\omega^2-\Delta_0^2}$. Our results for the LDOS therefore show maximum response at twice this wave vector length. Physically, this corresponds to scattering across the diameter of the contours of constant energy as is expected for such a circularly symmetric problem\cite{TamiQPI3}. This change in the singular value of $|\qv|$ with the frequency $\omega$ can be seen in the bottom plots of Fig.~\ref{fig:LDOSmuzero}. Here we plot the LDOS for several perturbations and probes. The singularities (or kink in the case of $\delta n_{11}$) clearly vary with $\omega$.

Next, we to study the effect of superconductivity on the QPI patterns. In short, the occurrence of $\Delta_0$ in the  critical radius of the LDOS described above is a signature of superconductivity. Without superconductivity the peak would occur at $2\omega$ and would thus persist all the way down to $\omega=0$\cite{GuoFranz}. With superconductivity present all features disappear once $\omega<\Delta_0$, {\it i.e.} when we probe energy scales within the gap. Thus having the pairing present in the system shifts the radius of this major feature to lower values.  We have explored this result in Fig.~\ref{fig:LDOSmuzero} by plotting several different LDOS patterns for varying $\omega$. For $\omega>\Delta_0$ we see that the singularity/kink in these patterns moves to smaller values of $|\qv|$ as $\omega$ decreases. For $\omega<\Delta_0$, or the solid curve in this figure, we see no signal at all.

Finally, we discuss the angular features of the QPI patterns we have calculated.  Studying Eq. (\ref{muzero}), we see that one requires the input $\alpha$ to be non-zero to find a pattern that is not circularly symmetric. In the case $\alpha=0$ we have either zero (for $\beta=1,2,3$) or a circularly symmetric function function when $\beta=0$.
Recall, the $\delta n_{00}$ result represents a non-magnetic impurity probed with a normal STM tip.
The other $\alpha=0$ results above would be those that are obtained from other perturbations and a normal STM tip. Thus our underlying {\it chiral} system does not show any angular dependence along the circular singularity in the QPI maps.  This is the case for a normal STM tip at $\mu=0$ regardless of the perturbation type.  That being said, as in Ref. [\onlinecite{GuoFranz}] we can see angular dependence in the $\alpha\ne0$ patterns, which corresponds to a spin-filtered STM tip. To look at these angular features we have presented density plots of our calculations on the lattice model in the top of Fig.~\ref{fig:LDOSmuzero}.

\subsection{General Chemical Potential}

\begin{figure*}[tb]
  \setlength{\unitlength}{1mm}
  \includegraphics[scale=.3571]{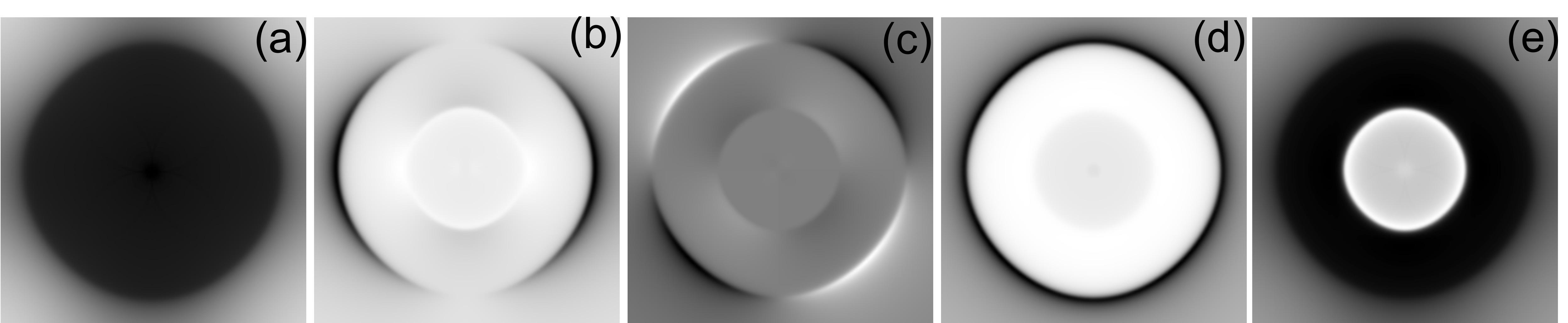}
\caption{{\small
Plots of the LDOS for the lattice model with $\mu\ne0$. Subfigures (a) through (e) label (respectively) results for $\delta n_{00}$, $\delta n_{11}$, $\delta n_{12}$, $\delta n_{33}$, and $\delta n_{OP}$ for the lattice problem . These calculations were done on a $600\times600$ lattice with $\lambda=1$, $t=0.5$, $\epsilon=2$, $\Delta_0=0.4$ and $\omega=0.6$, $\mu\ne0.5$
   }  }\label{fig:LDOSmufinite}
\end{figure*}

Moving to non-zero chemical potential makes calculating the LDOS exactly for the continuum model intractable. Therefore we must rely on numerics for both the continuum and lattice model. Our results for the $\mu\neq 0$ lattice model are presented in  Fig.~\ref{fig:LDOSmufinite}. These results lead us to make three observations. First, in the figure we notice that in addition to one major circular pattern, we can see a very subtle secondary pattern in the case of magnetic impurities. Second, all patterns essentially respect the same angular symmetry as at $\mu=0$. Third, the radius of the major pattern in the impurity QPI patterns changes noticeably when compared to the $\mu=0$ results whereas the OP suppression pattern does not.

Let us put these observations onto some more solid ground. We begin with the observation that the angular symmetry does not change. We have thus established that tuning the chemical potential away from the Dirac point does not lead to angular signatures in either the impurity or OP suppression perturbation. We now dedicate the rest of this subsection to explaining the other two observations above.

\begin{figure}[tb]
  \setlength{\unitlength}{1mm}
\begin{tabular}{cc}
   \includegraphics[scale=.35]{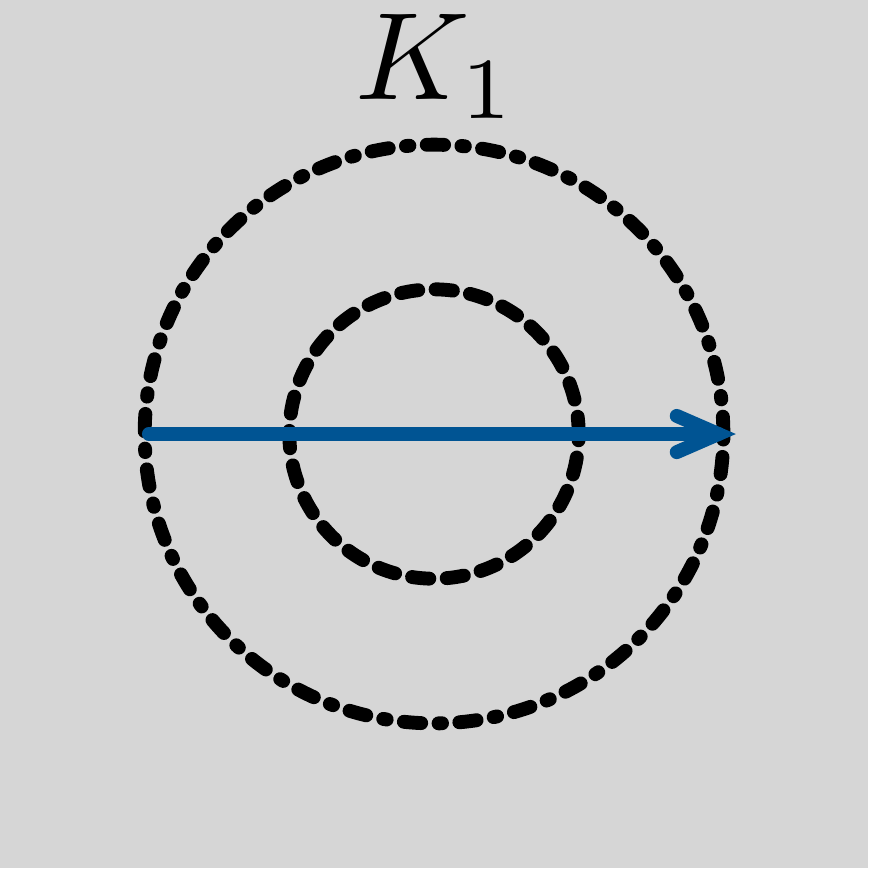} &
   \includegraphics[scale=.35]{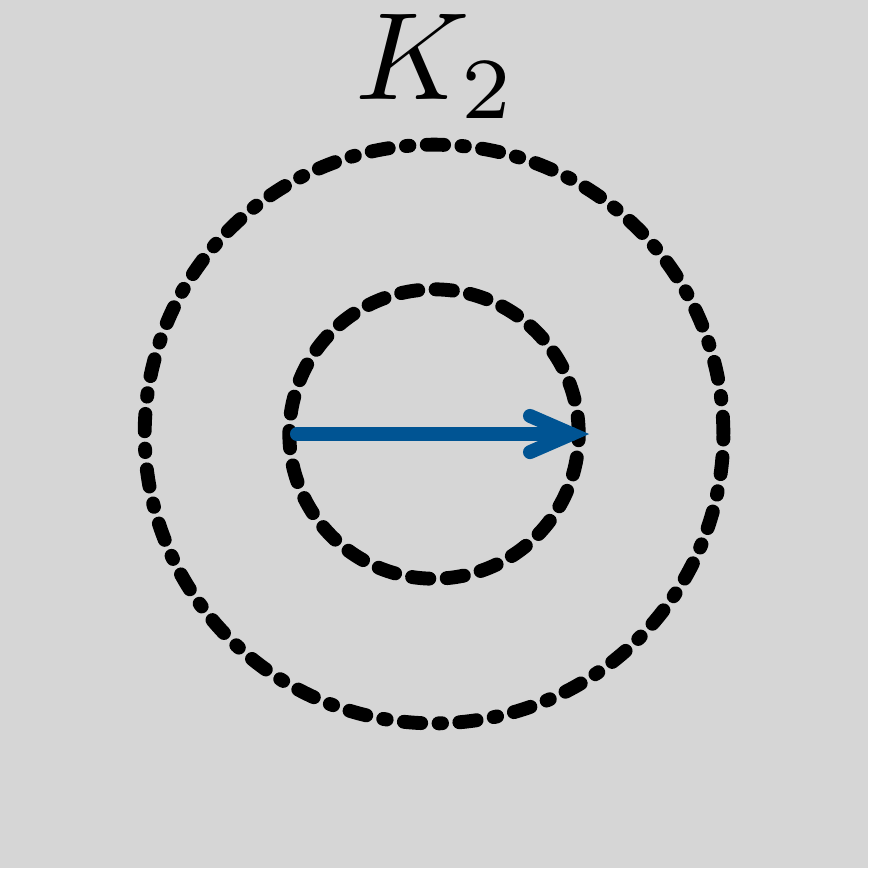} \\
   \includegraphics[scale=.35]{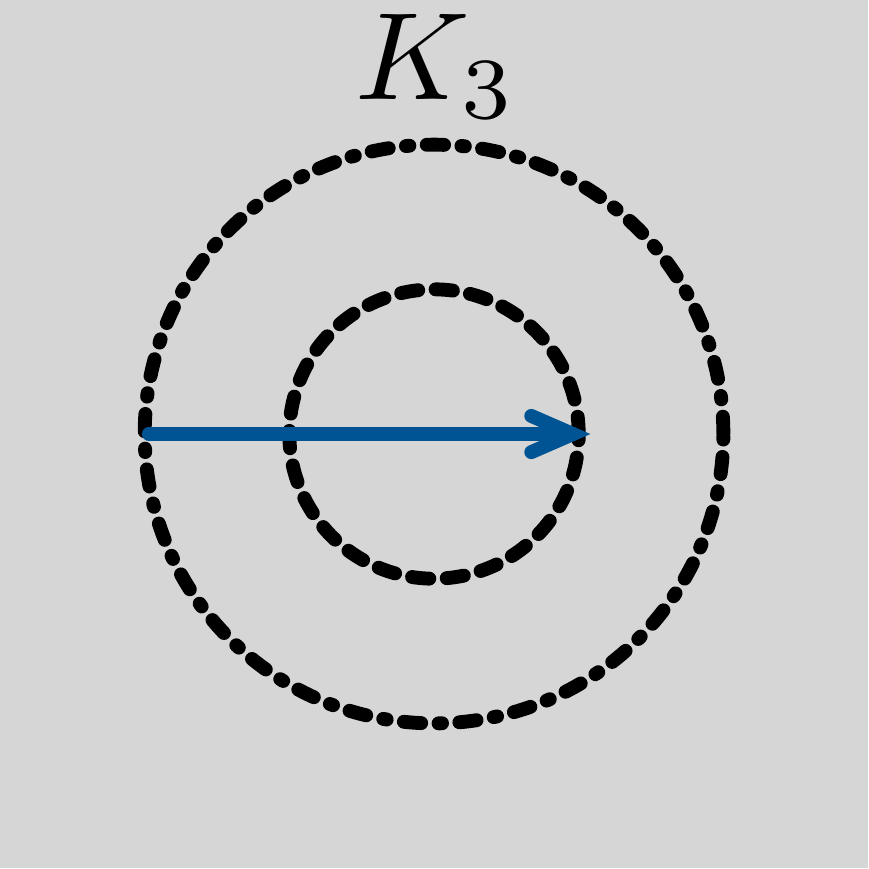} &
   \includegraphics[scale=.35]{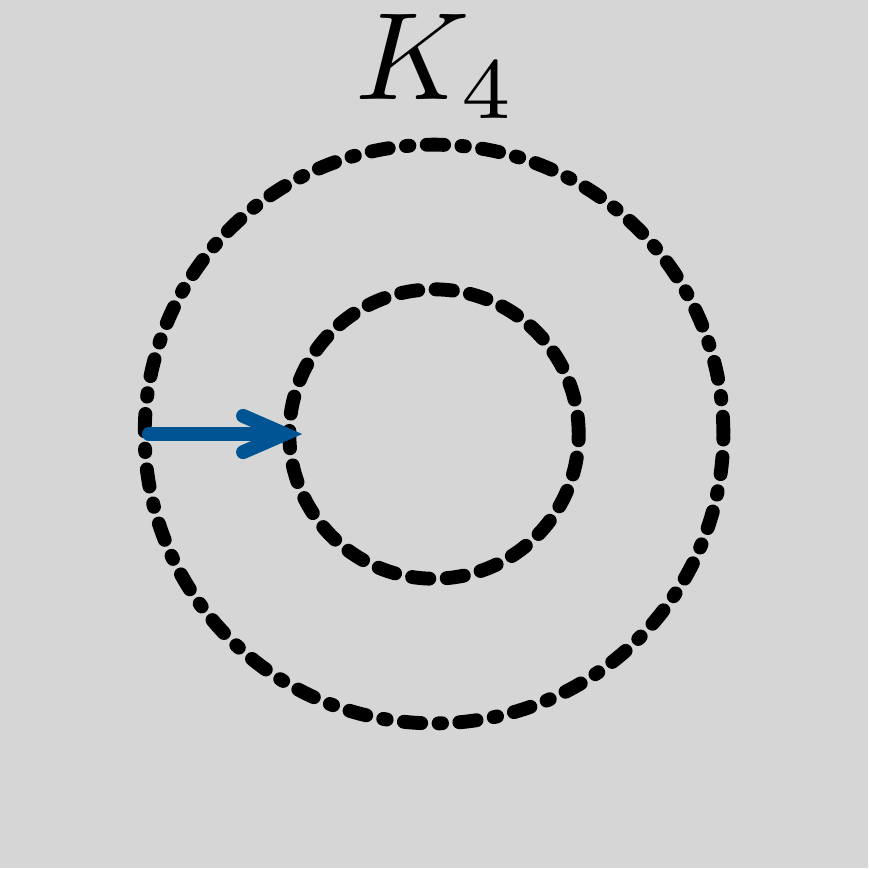}
   \end{tabular}

\caption{{\small
Various quasiparticle scattering processes which may contribute to the radius of the singularity in the LDOS. The broken lines represent the two contours of constant energy and the arrow shows the possible process.
     }
     }\label{fig:cartoon}
\end{figure}

To understand the above $\mu$ dependence, let us recall an observation from the last subsection; the major radial features in the LDOS patterns appear at values of $\qv$ corresponding to impurity scattering between states at the same quasiparticle energy. Additionally, this scattering occurs across the diameter of the contour of constant energy. At $\mu=0$ we have degenerate contours of constant energy. When $\mu$ is nonzero both of our models develop multiple contours of constant energy. In the continuum model, the radii of these contours are given by $|\kv|=|\mu\pm\sqrt{\omega^2-\Delta_0^2}|$. With these two contours one can imagine 4 quasiparticle scattering processes across the diameter of these circles. These processes are illustrated in Fig.~\ref{fig:cartoon} and labelled $K_1$ through $K_4$.

\begin{figure}[tb]
  \setlength{\unitlength}{1mm}
 \includegraphics[scale=.35]{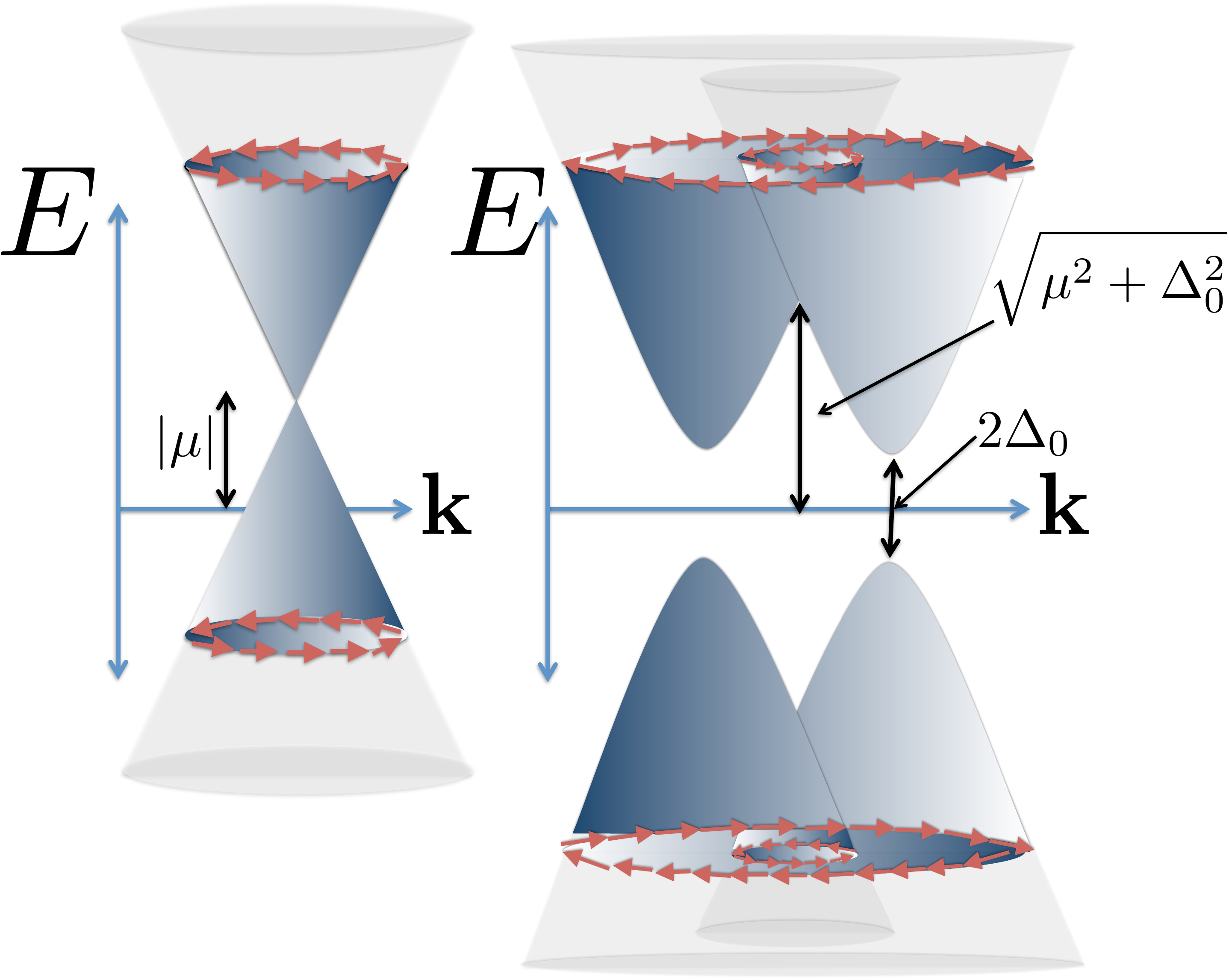}

\caption{{\small Dispersion with chirality indicated by arrows both before (left) and after (right) superconductivity is considered.
     }
     }\label{fig:DiracSC}
\end{figure}

With these different processes in mind we make the following two observations: (1) magnetic and non-magnetic impurities favor scattering processes between {\em the same } contour (intra-contour), (2) the OP suppression perturbation favors scattering processes between {\em different} contours (inter-contour). The QPI patterns of impurity scaterrers show a major singularity at $|\qv|=2K_1$ for $\mu>0$ with a minor singularity at $|\qv|=2K_2$. For $\mu<0$ the major singularity occurs at $|\qv|=2K_2$ with the secondary minor singularity at $|\qv|=2K_1$. Meanwhile, the pattern of an OP suppression shows singularities at $2K_3$ when $\mu<\sqrt{\omega^2-\Delta_0^2}$ and $2K_4$ when $\mu>\sqrt{\omega^2-\Delta_0^2}$

In relation to the above we find interesting mathematical relations for the radii of the singularities in the LDOS patterns. The impurity QPI patterns have their major singularity at $|\qv|=2\mu+2\sqrt{\omega^2-\Delta_0^2}\equiv q_{p}$ while the OP suppression pattern shows a singularity at $|\qv|=2\sqrt{\omega^2-\Delta_0^2}\equiv q_{a}$. Thus the intra-contour process increases linearly with $\mu$ while the inter-contour scattering wave vector length is {\em independent} of $\mu$.

We can understand the above results by heuristic arguments.  The two constant energy contours seen in the $\mu\neq 0$ case are the result of band reflection as depicted in Fig.~\ref{fig:DiracSC}.  In this figure the Dirac cone is intermitted by a gap and reflected about the $\epsilon_k = \mu$ horizontal line due to the introduction of superconductivity.  On each of these contours one can define a chirality angle $\phi_k = \arg(k_x+ik_y)$.  This chirality angle affects the wave function in two ways.  First, even without superconductivity, spin-orbit coupling locks the spin direction to the momentum direction.  Second, the pairing inherits this chirality angle, becoming effectively a $p$-wave superconductor.  Its order parameter winding appears in the coherence factors. The notion of chirality is depicted by arrows in  Fig.~\ref{fig:DiracSC}.

To describe this analytically we note that in our choice of basis the Hamiltonian has the form
\begin{equation}
\mathcal{H}_{\kv}=  \left( \begin{matrix} 
      -\mu & -vke^{i\phi_\kv}&0&\Delta_0 \\
      -vke^{-i\phi_\kv} & -\mu&-\Delta_0&0 \\
      0&-\Delta_0&\mu&-vke^{-i\phi_\kv}\\
      \Delta_0&0&-vke^{i\phi_\kv}&\mu\\
   \end{matrix}\right)
\end{equation}
and the positive energy eigenvectors have the form
\beq
|\psi_\pm(\kv)\rangle =  \left(  \begin{matrix} 
     \frac{ u_{k, \pm}}{\sqrt{2} }\left(  \begin{matrix} 
      \mp e^{i\phi_\kv} \\
      1  \\
   \end{matrix}\right)  \\
      \mp e^{i\phi_\kv} \frac{ v_{k, +}}{\sqrt{2} }\left(  \begin{matrix} 
      \pm e^{-i\phi_\kv} \\
      1  \\
   \end{matrix}\right)  \\
   \end{matrix}\right),\ E_{k,\pm}=\sqrt{\epsilon_{k,\pm}^2+\Delta^2}
\eeq
here $u_{k,\pm}=\sqrt{\frac{E_{k,\pm}+\epsilon_{k,\pm}}{2E_{k,\pm}}}$, $v_{k,\pm}=\sqrt{\frac{E_{k,\pm}-\epsilon_{k,\pm}}{2E_{k,\pm}}}$ and $\epsilon_{k,\pm}=-\mu\pm v|\kv|$.

The two branches of energy above (denoted with a $+$ or a $-$) have chirality directions which wind in {\em opposite} directions in the Brillouin zone (see Fig.~\ref{fig:DiracSC}). Notably, the chirality direction for a given branch is completely inverted upon sending $\kv\to-\kv$. This has important implications for our system where (as we have already discussed) scattering across the diameter of contours of constant energy is favored.

Looking at the impurity potentials and considering its action on $|\psi_\pm(\kv)\rangle$ one can see that it reverses the chirality angle. This means that $V^{\alpha}|\psi_\pm(\kv)\rangle$ will have the same chirality angle as $|\psi_\pm(-\kv)\rangle$. This explains the favoring of the $K_1$ and $K_2$ transitions.

Meanwhile, the OP suppression potential replaces both the spin and particle hole degree of freedom. As a result, the wavefunction is almost in tact (except for the Bogoliubov coherence factors being exchanged). This means $V^{OP}|\psi_\pm(\kv)\rangle$ will have the same chirality angle as $|\psi_\pm(\kv)\rangle$ which leads to a suppression of scattering across {\em the same} contour as here the initial state and the final state have {\em opposite} chirality angles.  Furthermore, these results lead to enhancement of inter-contour scattering. This is because there is a place on the other contour with the same chirality angle. For energies above the Dirac point we are cutting each branch only once and so the two contours we are left with have chirality angles winding in opposite directions. This leads to $K_3$ scattering processes. For energies below the Dirac point we cut the same energy branch twice and so we will have two contours with chirality angles winding in the same direction. For this case $K_4$ processes are favored.

\begin{figure}[tb]
  \setlength{\unitlength}{1mm}
\begin{tabular}{cc}
   \includegraphics[scale=.33]{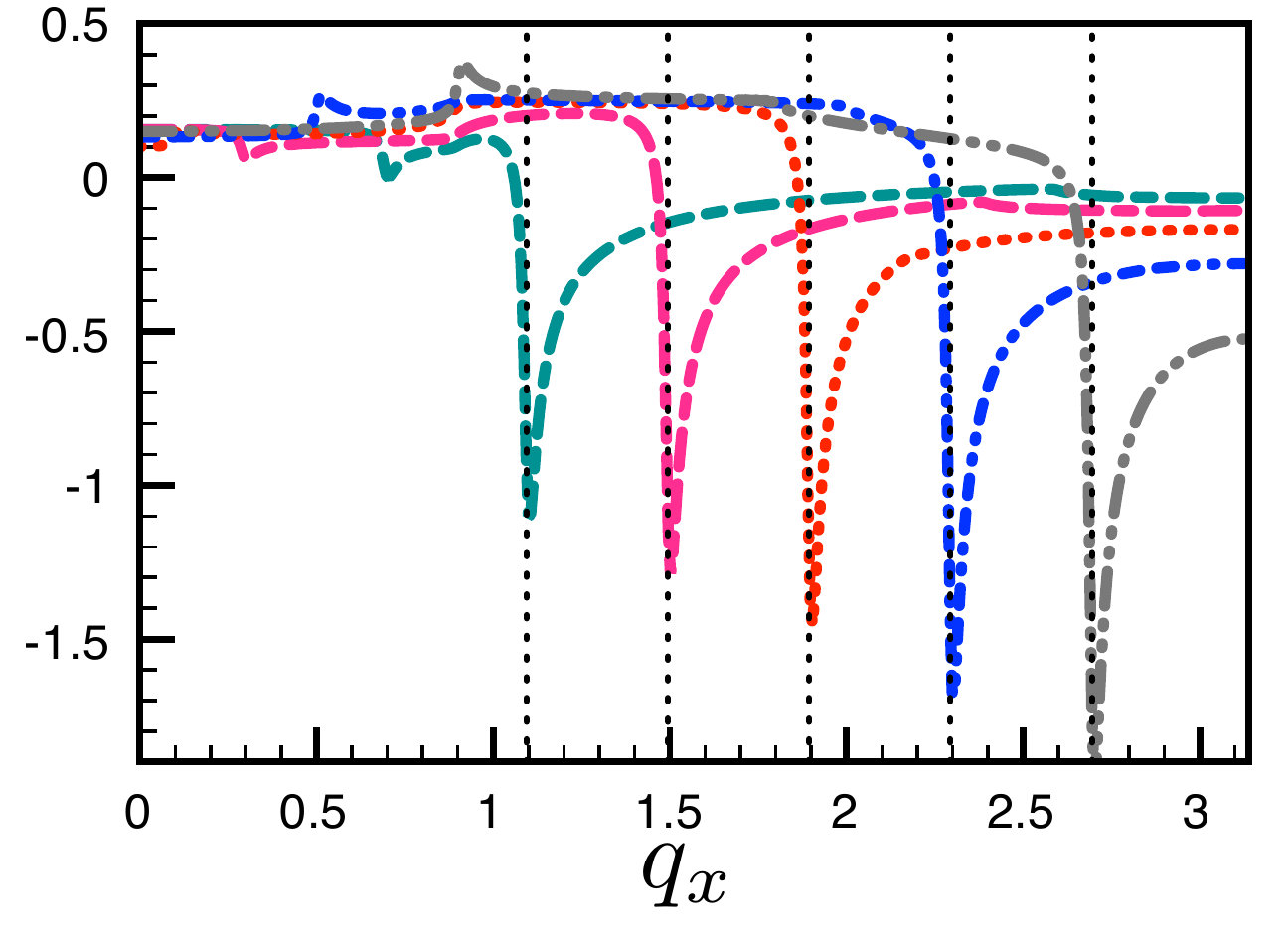}&  \includegraphics[scale=.33]{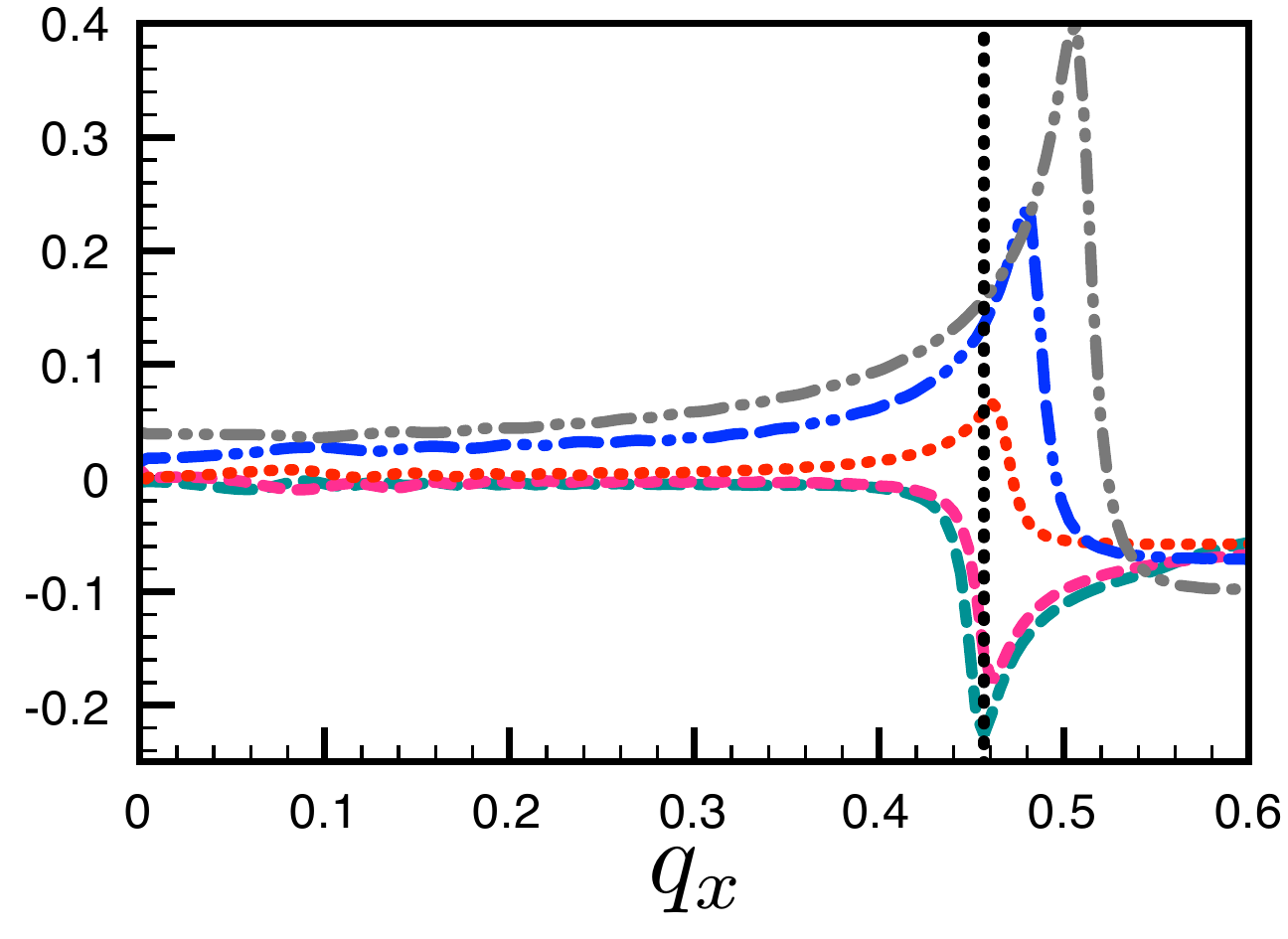}\\
   \includegraphics[scale=.33]{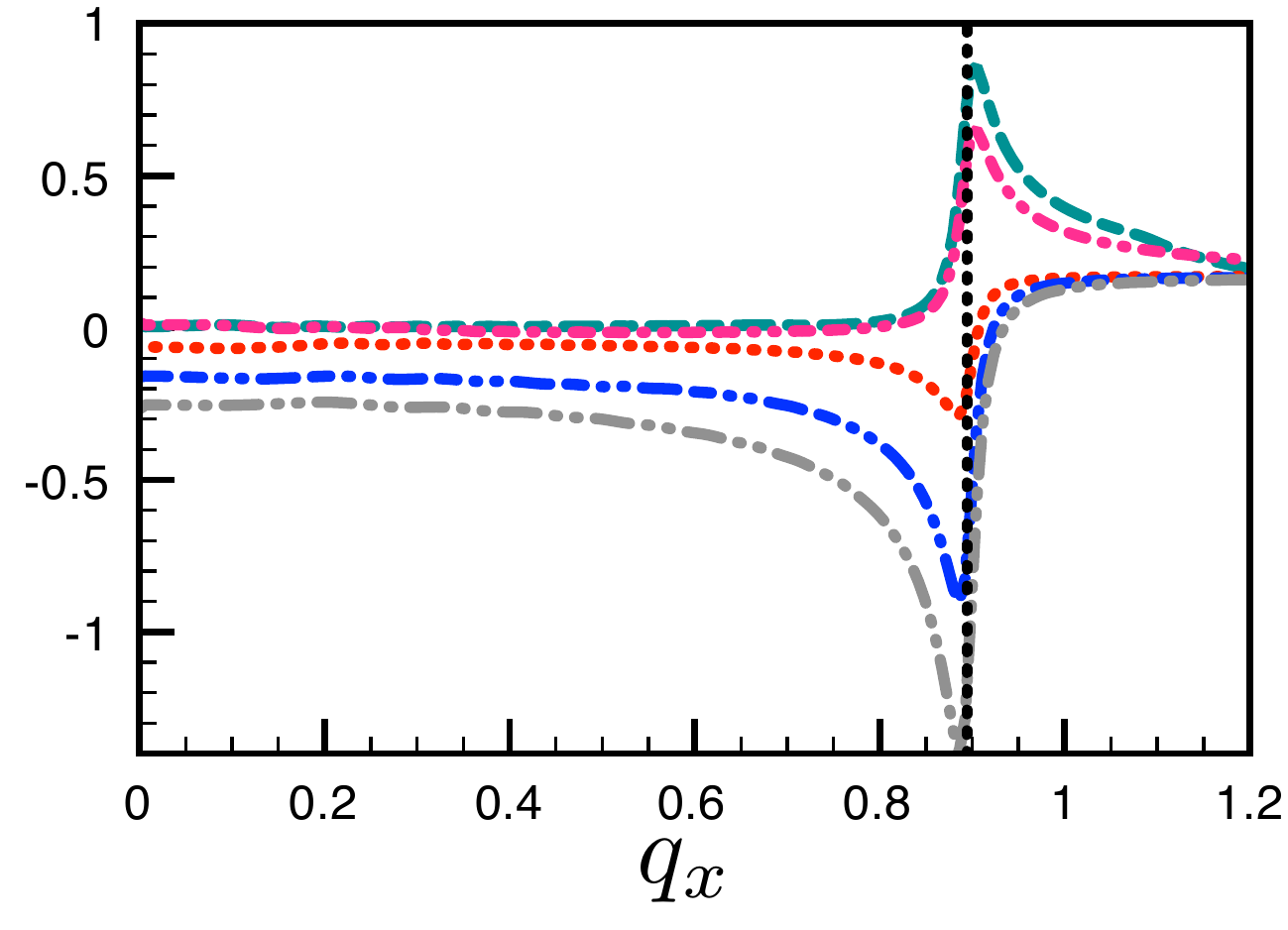} & \includegraphics[scale=.33]{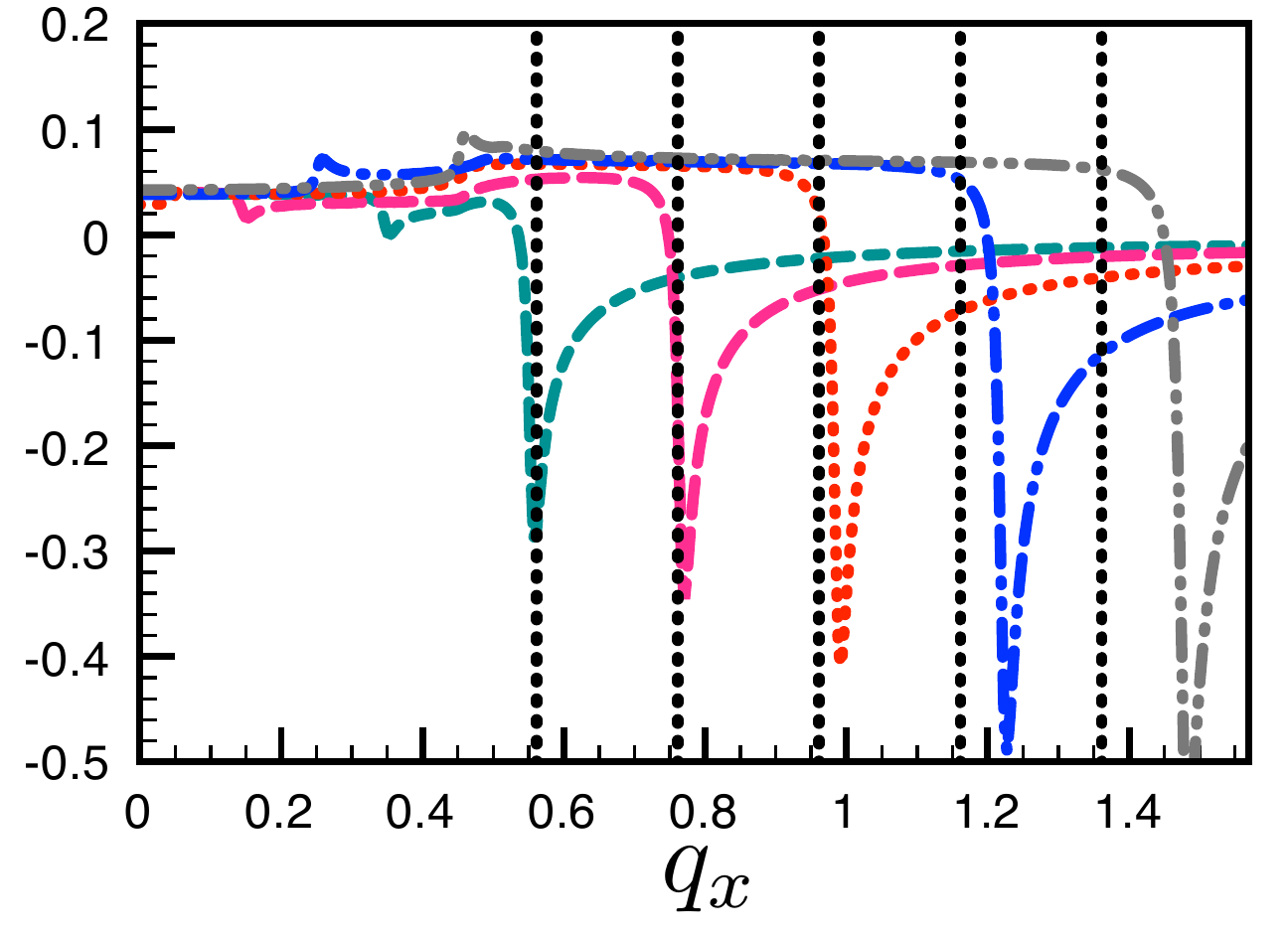}
   \end{tabular}
\includegraphics[scale=.4]{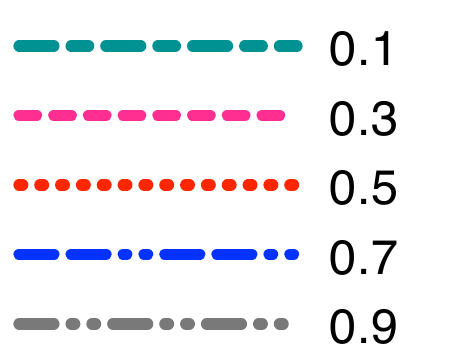}

\caption{{\small
Local density of states for finite chemical potential. On the top we show $\delta n_{3,3}$ along the $q_x$ axis for the continuum model (left) and the lattice model (right),  while on the bottom we show $\delta n$ for an anomalous impurity along the $q_x$ axis for the continuum model (left) and the lattice model (right). In both plots we have fixed $\omega=0.6$, $v=1$, $\Delta_0=.4$ and  the calculations were done on a $1500\times1500$ lattice of points. In each figure we have included vertical dashed lines at relevant $q_x$ values. In the top figures we have one at $2\mu +2\sqrt{\omega^2-\Delta_0^2}$ for each $\mu$ considered while at the bottom we have a single line at $2\sqrt{\omega^2-\Delta_0^2}$. On the bottom we have included a legend that labels the value of $\mu$ used to calculate the data in each curve.
     }
     }\label{fig:continuum_nonzero_mu}
\end{figure}

We explore the above results in our continuum model in the left of Fig.~\ref{fig:continuum_nonzero_mu}. Here we have plotted $\delta n_{33}$ and $\delta n_{OP}$ for several different values of the chemical potential. We see very clearly that the impurity scatterers ({\it e.g.} the $\delta n_{33}$ pattern) have singularities that occur at $|\qv|$ values that increase linearly with $\mu$. Meanwhile, the pattern for the OP suppression in this figure illustrates clearly that the singularity in this pattern does not depend on $\mu$.

We now contend that the functional form with respect to $\mu$ of $q_a=2\sqrt{\omega^2-\Delta_0^2}$, the radius for the OP suppression, and $q_p=2\mu+2\sqrt{\omega^2-\Delta_0^2}$, the radius for the regular impurity, ({\it i.e.} linear and constant respectively) is a result of the underlying Dirac band structure. To argue this we approximate the spectrum by $E=\sqrt{\Delta_0^2+(\tilde{\epsilon}_{\kv}-\mu)^2}$ where $\tilde{\epsilon}_{\kv}$ is the dispersion of the underlying band structure. If we wish to invert this equation to find $\kv$ as a function of $E$, i.e. to find the equation for the contours of constant energy, we obtain $\tilde{\epsilon}_\kv = \mu\pm\sqrt{E^2-\Delta_0^2}$.

Now let us think about a circularly symmetric dispersion for simplicity. Then $\tilde{\epsilon}_\kv =\tilde{\epsilon}_k$. Let us assume that we are close enough to the center of the Brillouin zone so that the leading order term in an expansion of $\tilde{\epsilon}_k$ is valid. We then take $\tilde{\epsilon}_\kv=\epsilon k^\gamma$ where $\gamma$ is some number and $\epsilon$ is a constant. Then the circles of constant energy  have a radius given by
\beq
k_E = \left(\frac{\mu\pm\sqrt{E^2-\Delta_0^2}}{\epsilon}\right)^{1/\gamma}
\eeq
We then see that, provided $\sqrt{E^2-\Delta_0^2}$ and $\mu$ are of comparable size, the only type of dispersion that yields energy contours that depend on $\mu$ in a linear matter is $\gamma=1$, or a Dirac-like dispersion. Thus this linear scaling of the major singularity in the QPI patterns is a property of a linearly dispersing band structure. Furthermore, the independence on $\mu$ of scattering from one contour to another is also a consequence of having a linear dispersion. The radius of singularities for such a process will be
\beq
K= \left(\frac{\mu+\sqrt{E^2-\Delta_0^2}}{\epsilon}\right)^{1/\gamma}-\left(\frac{\mu-\sqrt{E^2-\Delta_0^2}}{\epsilon}\right)^{1/\gamma}
\eeq
The above is $\mu$ independent for $\gamma=1$ only.

The above argument of course only holds provided the $k$ values we are interested in are suitably close to the $\Gamma$-point (or, more generally, wherever the Dirac point occurs) so that our linear approximation is valid. That is to say, in a realistic system $\tilde{\epsilon}_\kv$ will contain other subleading contributions on top of the linear Dirac like term. To explore how well our results hold in such a system we have calculated $\delta n_{33}$ and $\delta n_{OP}$ for several different values of the chemical potential in our lattice model. These results are presented on the right of Fig.~\ref{fig:continuum_nonzero_mu}. In this figure we see that our observation holds very well provided that $\mu$ is kept small enough. For intermediate values of $\mu$ deviations increase as $\kv$ moves away from the Dirac point.

\subsection{System in a Capacitor}

\begin{figure}[tb]
  \setlength{\unitlength}{1mm}
\begin{tabular}{cc}
   \includegraphics[scale=.33]{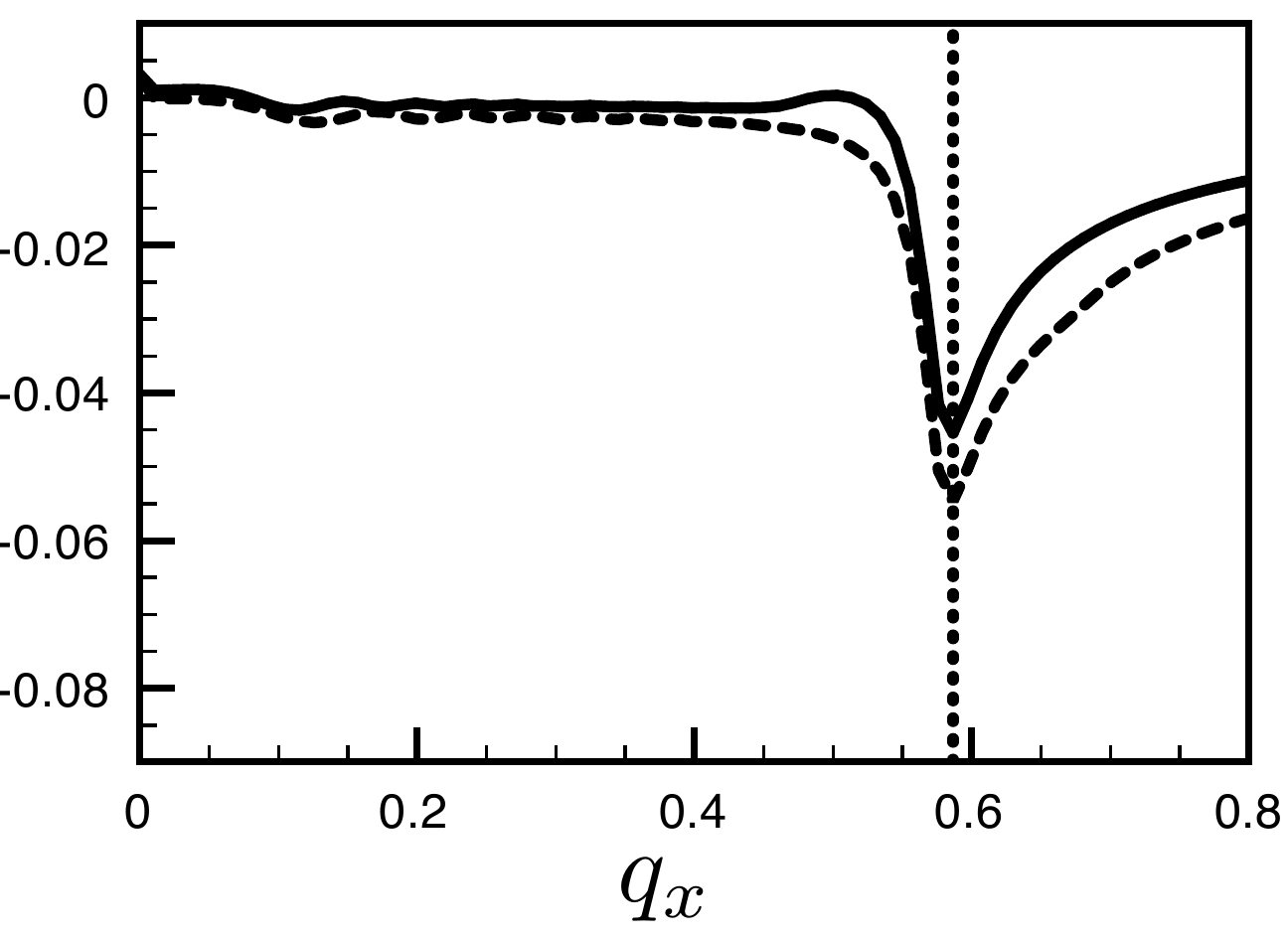}&  \includegraphics[scale=.33]{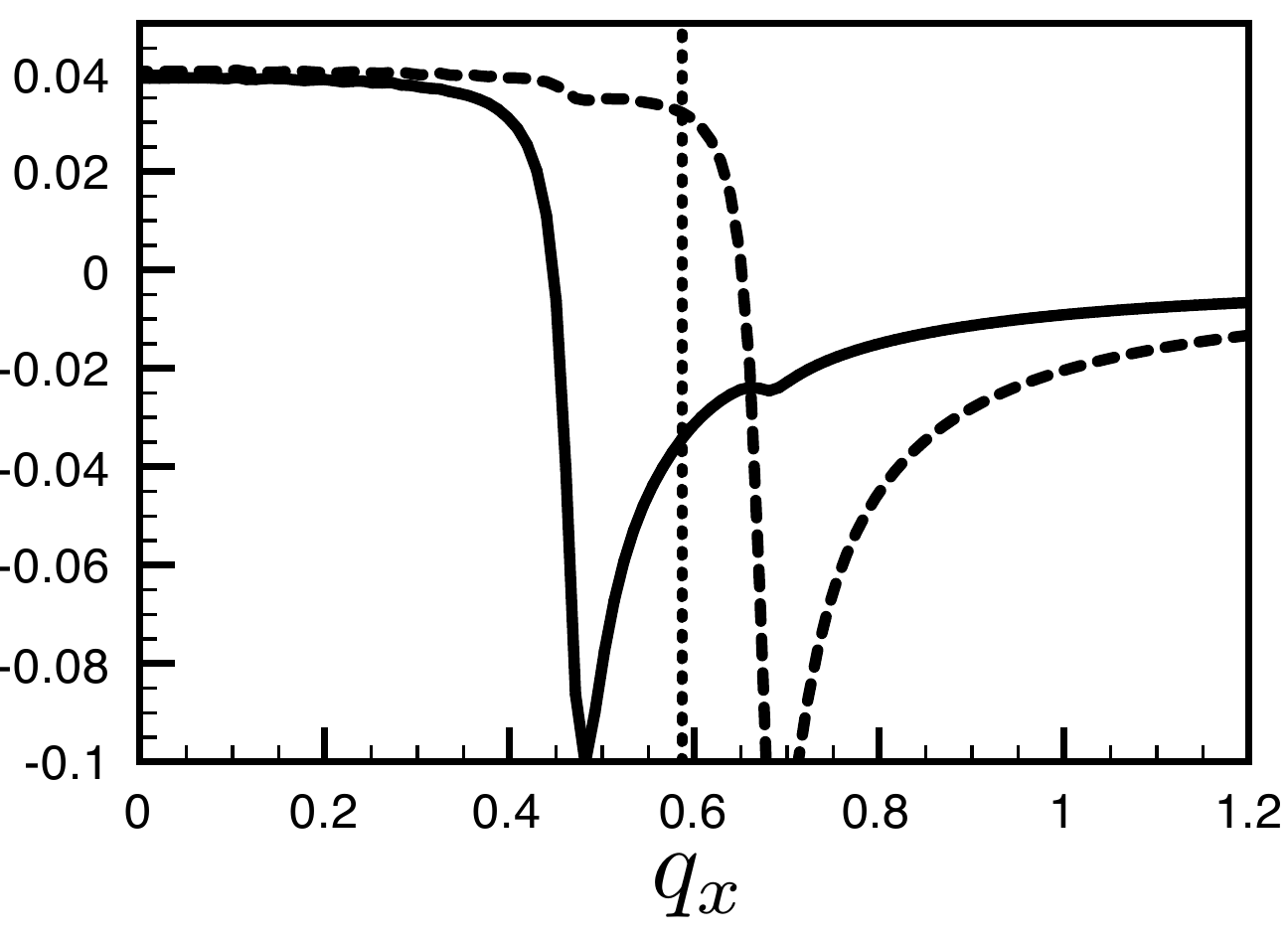}\\
   \includegraphics[scale=.33]{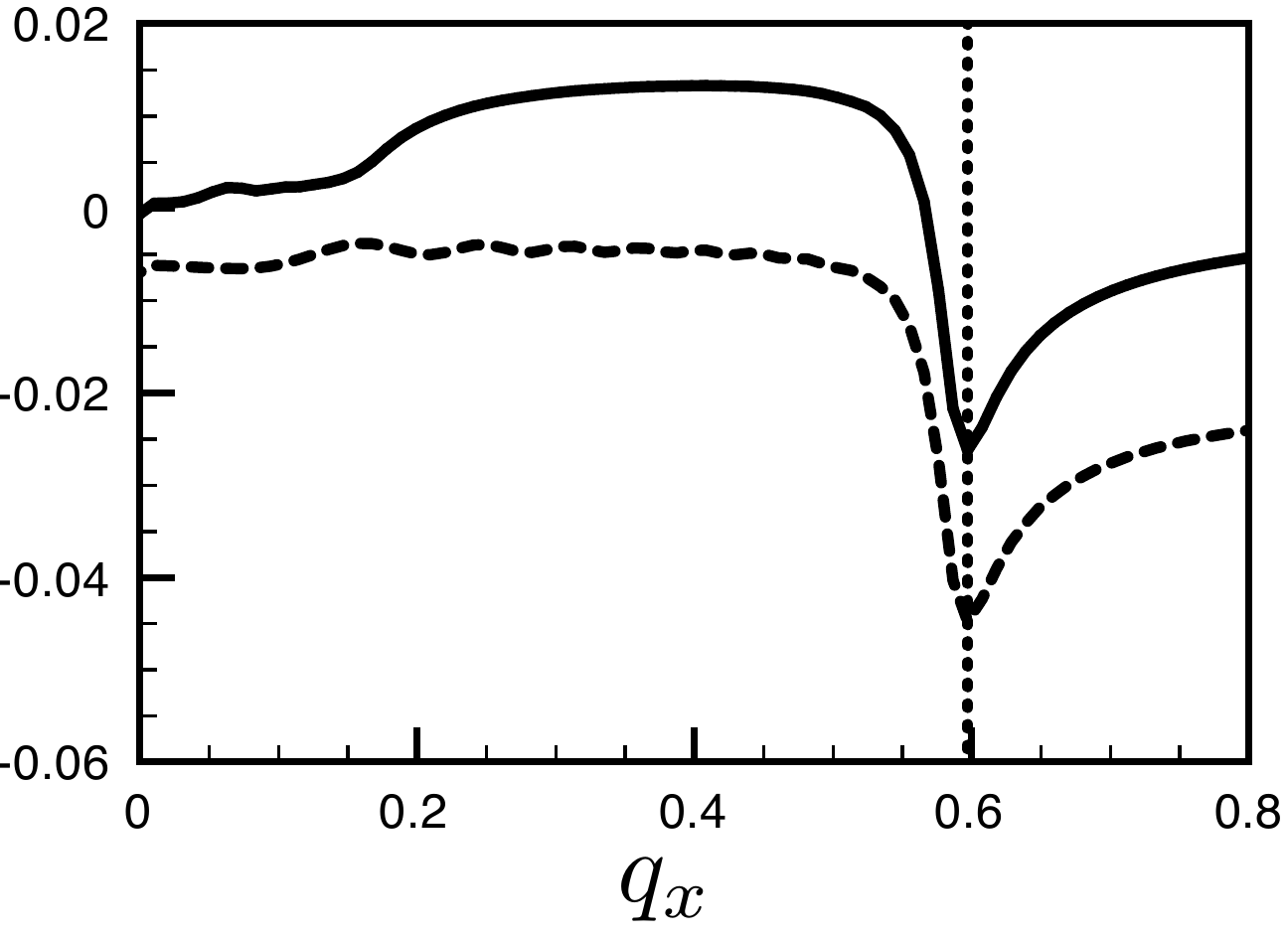} & \includegraphics[scale=.33]{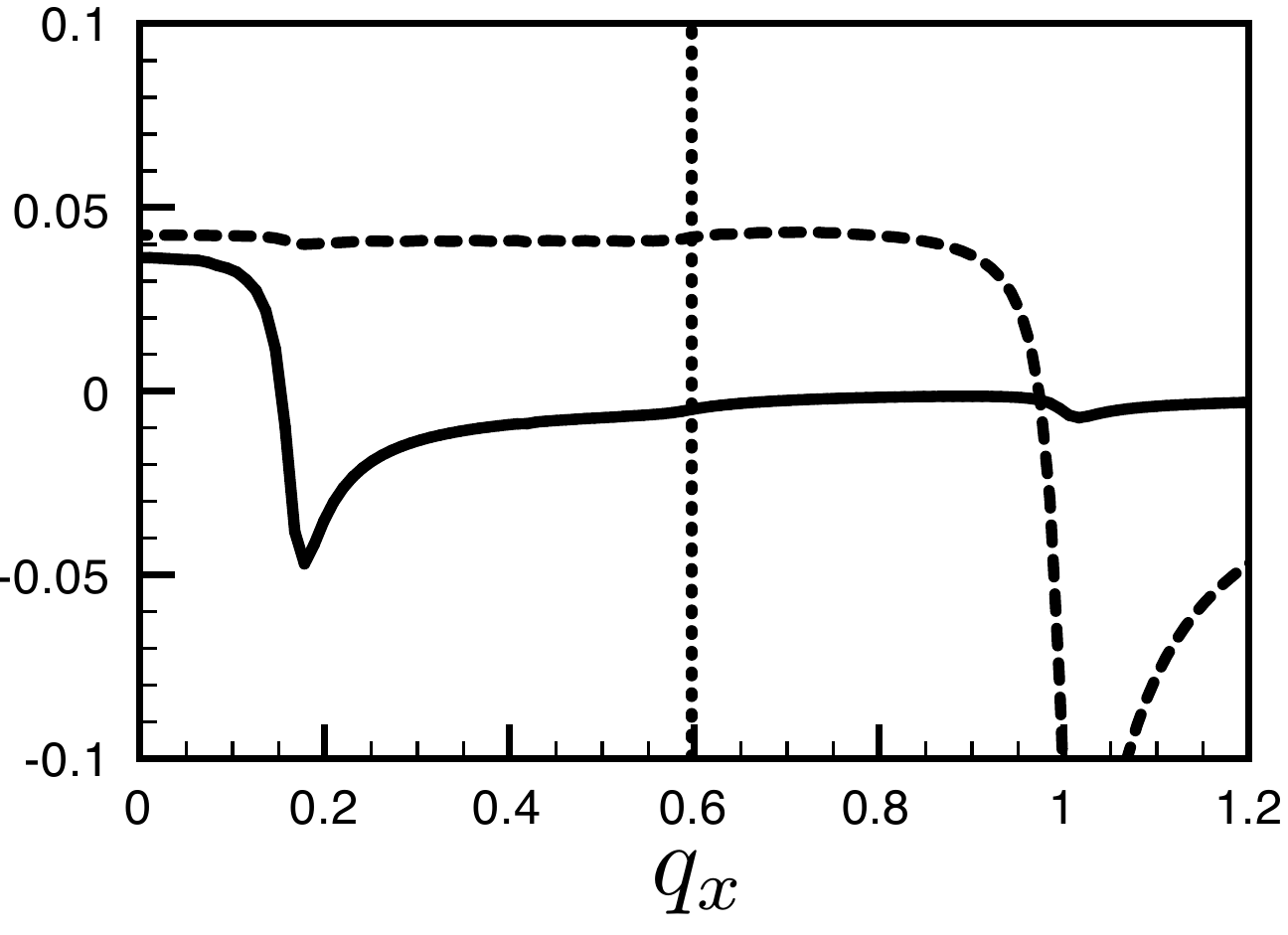}
   \end{tabular}
\caption{{\small
Local density of states for a system in a capacitor. In each figure the solid line is the QPI pattern on the top of the sample while the broken line is the pattern on the bottom of the sample. Along the top we have plotted $\delta n_{OP}$ (left) and $\delta n_{33}$ (right) for $V=0.1$ while along the bottom we have $\delta n_{OP}$ (left) and $\delta n_{33}$ (right) for $V=0.4$.  In both plots we have fixed $\omega=0.6$, $v=1$, $\Delta_0=.4$, $\mu=0$ and the calculations were done on a $600\times600$ lattice of points. The plots show cuts of the QPI pattern along the line $q_y=0$. The vertical dashed line in each figure is a guide to show where the anomalous pattern is peaked.
     }
     }\label{fig:lattice_capacitor}
\end{figure}

In the previous subsection we have outlined an interesting dependence of the radii of the major feature in the LDOS on the chemical potential. The chemical potential can be tuned in STIs by chemical doping or, in a thin flake, by electrostatic gating.  To illustrate the effect of a non-zero chemical potential in a simple setting we imagine placing a system inside a capacitor which has an effect of biasing the system so that there is a potential $V$ at the top and $-V$ at the bottom. As mentioned in the introduction of this work such a scenario could be realized through the use of dual-gate structures\cite{ChangChemicalPotential}.  We describe this by adding
\beq
H_{Cap} =  V \text{diag}(\sigma_0, \sigma_0, -\sigma_0,- \sigma_0, -\sigma_0, -\sigma_0, \sigma_0, \sigma_0)
\eeq
to our clean Hamiltonian in Eq. (\ref{lattice}).

In our lattice model the two surfaces (top and bottom) are only weakly coupled and so we may think of this as introducing a chemical potential to the two surfaces. This chemical potential is equal in magnitude but opposite in sign on each edge. Let us, without loss of generality, assume the positive bias is introduced on top of the sample (notice that $H_{Cap}$ enters the Hamiltonian different by a minus sign from the chemical potential). Thinking of the two edges as isolated Dirac points, our findings thus far in the paper then simply predict the following: The radius of the major peak in the LDOS pattern for an impurity QPI should decrease on the top of the sample (as if $\mu=-V$), and decrease on the bottom (as if $\mu=V$). Meanwhile, an OP suppression should look the same on the top and bottom of the sample.

In addition to the qualitative prediction made above, the strength of the bias in this setup is, at least in principle, tuneable {\em via} external means.

In Fig~\ref{fig:lattice_capacitor} we present the LDOS on the top and bottom of a sample for several values of this bias voltage.  We see that the exact scenario described above plays out, namely as $V$ is increased the singularity in the impurity QPI pattern moves to larger $|\qv|$ on one edge and smaller $|\qv|$ on the other. In addition, we see that the OP suppression placed on the top and bottom of the sample give identical results and are essentially independent of this bias voltage.

\section{Conclusions}

We have studied the QPI patterns induced by different local perturbations in both a lattice and a continuum model for a SC surface of a three dimensional topological insulator. Our results for a half filled band (such that the chemical potential is at the Dirac point) are similar to those calculated in the past for a strong topological insulator surface. The existence of superconductivity in the system gaps out the low-lying excitations and shows up qualitatively in the radius of the singularity/kink that occurs in the QPI pattern. The presence of $\Delta_0$ reduces the critical radius of this singularity. For STM bias values below the superconductive gap no signal can be observed. For non-magnetic impurities the QPI response is weak and consists of an edge similar to the normal state result. Remarkably though, we find that when disorder in the order parameter amplitude is included (one that will generically be present) the edge transforms into a peak, which should be more easily observed in experiment. The shape of the singularity -- edge vs. peak --
can thus be used as an indicator of the dominant source of quasiparticle scattering in the sample.

With a finite chemical potential we find almost no change in the angular features of the QPI pattern. We do however find that the quasiparticle scattering processes contributing to impurity and OP suppression scattering are different. As a result, the singular features of the impurity QPI pattern depend linearly on the chemical potential and that those of the OP suppression are independent of the chemical potential. We argued that this functional dependence is unique to a Dirac-like spectrum and showed that it approximately holds in our more complicated lattice model. We have also proposed and verified with a calculation an alternative method to tune the chemical potential by placing the sample in a capacitor.

\section{Acknowledgements}
 The authors are thankful for useful discussions with J. Hoffman Financial support for this work was provided by the NSERC and FQRNT (TPB), the Vanier Canada Graduate Scholarship (AF) and CIfAR. Some of the numerical calculations for this work were performed using CLUMEQ/McGill HPC supercomputing resources.

\bibliographystyle{apsrev}
\bibliography{QPI}

\appendix

\section{Impurity Potentials}
Here we will describe the mathematical expressions we have used to describe the impurity potential. Starting with the charge impurity we define $V^{charge}_{imp}=V^{0}=\mathcal{I} \otimes \tau_z$ where $\mathcal{I}= \sigma^{0}\otimes \tilde{\mathcal{I}}$ acts on spin and any other degree of freedom in the system and $\tau$ acts on Nambu space. The operator $\tilde{\mathcal{I}}$ acts, in a suitable manner, on any other degrees of freedom that may be present ({\it i.e.} not spin or particle-hole). For example, in our lattice model $ \tilde{\mathcal{I}} _{\alpha,\alpha'}$ acts on the top-bottom surface degree of freedom while in the continuum model there are no other degrees of freedom left. In the lattice model we consider an impurity localized on a particular edge and so  $ \tilde{\mathcal{I}} _{\alpha,\alpha'}$  is diagonal with entries of $1$ on the impurity edge and $0$ on the other edge.

Secondly, we are interested in magnetic impurities. These alter the local Zeeman splitting on a single site and so couple to the spin of the electrons.  We consider three separate impurities (one for each cartesian direction) while noting that a general magnetic impurity can be written as a linear combination of these potentials. As such we define (in Nambu space) ${V}^{\beta\ne0}=\text{diag}\left(\tilde{V}^{\beta\ne0} ,-(\tilde{V}^{\beta\ne0})^* \right)$ where $\tilde{V}^{\beta\ne0}=\sigma^{\beta}\otimes \tilde{\mathcal{I}} $.

Finally, we consider our treatment of the OP suppression perturbation. This is modelled by taking  ${V}^{OP}=(i\sigma^y \otimes \tilde{\mathcal{I}})\otimes (i\tau^y)$ where we remind the reader that $\sigma^i$ act on spin degrees of freedom, $\tau^i$ on Nambu space and $ \tilde{\mathcal{I}}$ is as defined above. We calculate the OP QPI pattern by using this potential in Eq.~(\ref{LDOScharge}).

Let us close this subsection with a brief description of the role of the probe potential, $V_\alpha$. In the main text we have imagined an experimental set-up where the physical STM tip is capable of resolving the component of the electron spin along a particular projection. In this case we are interested in only certain components of the change in the Greens function matrix. To find these relevant components and how they contribute to an interference pattern we must place an operator in the trace in Eq.~(\ref{LDOScharge}) which acts to find the proper contribution. For example, if we are interested in an STM tip which resolves the $z$-component of the electron spin we would add a $\sigma^z$ to Eq.~(\ref{LDOScharge}). In general, we call this matrix $V^\alpha$ where $V^\alpha$ is defined in the same way as $V^{\beta}$ above. For $\alpha=0$ we are considering only a normal STM tip, while for $\alpha=1,2,3$ we consider an STM tip capable of resolving the spin in the $x$, $y$, or $z$ direction respectively. In either case we additionally consider the physically relevant case of the tip only resolving degrees of freedom on one edge of the system; the tip can only make contact with one edge or the other. 

\section{Exact Green's Function for Continuum Model}

Here we will outline our calculation of the exact, clean Green's function for the simple Dirac continuum model. To find the Green function of $H_0$ we begin by diagonalizing $H_{TI}$. This is accomplished by making the transformation $c_{\kv} = \bar{U}_{\kv} b_{\kv}$ where
\beq
\bar{U}_{\kv}=\frac{1}{\sqrt{2}} \left(   \begin{matrix} 
      -e^{i\phi_{\kv}} &e^{i\phi_{\kv}} \\
      1& 1 \\
   \end{matrix}\right)
\eeq
where $\phi_{\kv}$ is the phase of $k_y+ik_x$ . In the above the operators $b_{\kv, \uparrow}$ ($b_{\kv, \uparrow}$ ) annihilate electrons from the upper (lower) Dirac cone. Using this transformation we can rewrite $\psi_{\kv} = T_{\kv} \eta_{\kv}$ where $\eta_{\kv} = ( b_{\kv},  b_{-\kv}^\dagger)^T$ and
\beq
T_{\kv}=\left(   \begin{matrix} 
      \bar{U}_{\kv}&0\\
      0& \bar{U}^*_{-\kv} \\
   \end{matrix}\right)
\eeq
In terms of this transformation the Green's function can be written
\beq
G(\kv, i\omega_m) = T_{\kv} \hat{G}(\kv, i\omega_m)T^\dagger_{\kv}
\eeq
where $\hat{G}(\kv, i\omega_m)$ is the Green function in the $\eta_{\kv} $ basis. After making this change of basis we can write $H_0$ as
\beq
H_0 =\frac{1}{2} \sum_{\kv} \eta_{\kv}^\dagger \hat{\mathcal{H}}_{\kv} \eta_{\kv}
\eeq
where
\beq
\hat{\mathcal{H}}_{\kv} =\left(   \begin{matrix} 
      \epsilon_{\kv,+} & 0 &  -\Delta_0 e^{-i\phi_{\kv}} & 0 \\
      0 &  \epsilon_{\kv,-} &  0 &  \Delta_0 e^{-i\phi_{\kv}}  \\
          -\Delta_0 e^{i\phi_{\kv}} & 0&  - \epsilon_{\kv,+}  & 0 \\
        0 &  \Delta_0 e^{-i\phi_{\kv}} &  0 & - \epsilon_{\kv,-}  \\
   \end{matrix}\right)
\eeq
where $\epsilon_{\kv,\pm} = -\mu \pm vk$. We see from the above matrix that the two bands are completely decoupled from each other and we have $p$-wave pairing on each Dirac cone. We have the two independent systems
\beqa
&&\hat{\mathcal{H}}_{\kv, +} = \left(   \begin{matrix} 
      \epsilon_{\kv,+}  &  -\Delta_0 e^{-i\phi_{\kv}}  \\
          -\Delta_0 e^{i\phi_{\kv}} &  - \epsilon_{\kv,+}   \\
   \end{matrix}\right)
  \\ \nonumber &&
   \hat{\mathcal{H}}_{\kv, -} = \left(   \begin{matrix} 
      \epsilon_{\kv,-}  &  \Delta_0 e^{-i\phi_{\kv}}  \\
          \Delta_0 e^{i\phi_{\kv}} &  - \epsilon_{\kv,-}   \\
   \end{matrix}\right)
\eeqa
Defining the two Green's function $(i\omega_m - \hat{\mathcal{H}}_{\kv, \lambda} )\hat{G}_{\kv, \lambda} =1$ where $\lambda =\pm1$ it is straightforward to show
\beq
\hat{G}_{\kv, \lambda} = \frac{1}{(i\omega_m)^2-E_{\kv,\lambda}^2}\left(   \begin{matrix} 
     i\omega_m+  \epsilon_{\kv,\lambda}  &  -\lambda \Delta_0 e^{-i\phi_{\kv}}  \\
         -\lambda \Delta_0 e^{i\phi_{\kv}} &  i\omega_m- \epsilon_{\kv,\lambda}   \\
   \end{matrix}\right)
\eeq
where $E_{\kv,\pm} = \sqrt{\epsilon_{\kv,\pm}^2+\Delta_0^2}$. Our Full Green function is then
\beq
\hat{G}_{\kv}(i\omega_m)  = \left(   \begin{matrix} 
      \hat{ g}_{\kv}(i\omega_m) &- \Delta_0 e^{-i\phi_{\kv}}  \hat{ f}_{\kv}(i\omega_m)  \\
       -\Delta_0 e^{i\phi_{\kv}}  \hat{ f}_{\kv}(i\omega_m)  &- \hat{ g}^*_{-\kv}(i\omega_m)  \\
   \end{matrix}\right)
\eeq
where
\beqa
&&\hat{ g}_{\kv}(i\omega_m) =\left(   \begin{matrix} 
      \frac{i\omega_m+\epsilon_{\kv,+}}{(i\omega_m)^2-E_{\kv,+}^2}   & 0 \\
       0&  \frac{i\omega_m+\epsilon_{\kv,-}}{(i\omega_m)^2-E_{\kv,-}^2} \\
    \end{matrix}\right)  \\ \nonumber &&
\hat{ f}_{\kv}(i\omega_m) =\left(   \begin{matrix} 
      \frac{1}{(i\omega_m)^2-E_{\kv,+}^2}   & 0 \\
       0& - \frac{1}{(i\omega_m)^2-E_{\kv,-}^2} \\
    \end{matrix}\right)
\eeqa
Now we apply the matrices $T_{\kv}$ to get back to the Green's function in a spin basis. It reads
\beq
{G}_{\kv}(i\omega_m)  = \left(   \begin{matrix} 
      { g}_{\kv}(i\omega_m) & { f}_{\kv}(i\omega_m)  \\
       { f}^\dagger _{\kv}(i\omega_m)  &- { g}^*_{-\kv}(i\omega_m)  \\
   \end{matrix}\right)
\eeq
where $ { g}_{\kv}(i\omega_m)= \bar{U}_{\kv} \hat{g}_{\kv}(i\omega_m)  \bar{U}_{\kv} ^\dagger $ and ${ f}_{\kv}(i\omega_m)  = \bar{U}_{\kv} \hat{g}_{\kv}(i\omega_m)  \bar{U}_{-\kv} ^T $. Performing the matrix multiplication one can show that
\begin{eqnarray}
&& { g}_{\kv}(i\omega_m) \\ \nonumber &&\frac{1}{2}\small{\left(   \begin{matrix} 
     g_{\kv,+}  +   g_{\kv,-} & e^{i\phi_{\kv}} \left( g_{\kv,-}-g_{\kv,+}  \right) \\
       e^{-i\phi_{\kv}} \left( g_{\kv,-}-g_{\kv,+}  \right) &  g_{\kv,+}  +   g_{\kv,-}  \\
    \end{matrix}\right) }
\end{eqnarray}
and
\begin{eqnarray}
&&{ f}_{\kv}(i\omega_m) = \\ \nonumber &&\frac{1}{2} \small{\left(   \begin{matrix} 
    e^{i\phi_{\kv}}   (f_{\kv,+}  -   f_{\kv,-} )&  \left( f_{\kv,-}+f_{\kv,+}  \right) \\
      - \left( f_{\kv,-}+f_{\kv,+}  \right) &  -e^{-i\phi_{\kv}}  (f_{\kv,+}  -   f_{\kv,-}  )\\
    \end{matrix}\right)}
\end{eqnarray}

where $g_{\kv,\lambda} = \frac{i\omega_m+ \epsilon_{\kv,\lambda}}{(i\omega_m)^2-E_{\kv,\lambda}^2}$ and $f_{\kv,\lambda}=\frac{\Delta_0}{(i\omega_m)^2-E_{\kv,\lambda}^2}$.

\end{document}